\documentclass[superscriptaddress,aps,prd,preprintnumbers,abbrv,nofootinbib,twocolumn]{revtex4-1}

\usepackage{amsfonts,amsmath,amssymb}
\usepackage{bm,tensor}
\usepackage{graphicx}
\usepackage{braket}
\usepackage{ifpdf}
\usepackage[utf8]{inputenc}
\usepackage{color}
\usepackage{xfrac}
\usepackage{comment}
\usepackage[colorlinks=true
,urlcolor=blue
,anchorcolor=blue
,citecolor=blue
,filecolor=blue
,linkcolor=blue
,menucolor=blue
,linktocpage=true
,breaklinks=true
,pdfproducer=medialab
,pdfa=true
]{hyperref}
\usepackage[normalem]{ulem} %to strike the words

\graphicspath{{./fig/}{./png/}}
\newcommand{\EQ}{\begin{equation}}
\newcommand{\EN}{\end{equation}}
\newcommand{\EQA}{\begin{eqnarray}}
\newcommand{\ENA}{\end{eqnarray}}

\newcommand{\Eq}[1]{Eq.~(\ref{#1})}
\newcommand{\Eqs}[2]{Eqs.~(\ref{#1}) and~(\ref{#2})}

\newcommand{\App}[1]{Appendix~\ref{#1}}

\newcommand{\Sec}[1]{Sect.~\ref{#1}}

\newcommand{\Fig}[1]{Fig.~\ref{#1}}
\newcommand{\FFig}[1]{Figure~\ref{#1}}

\newcommand{\Figsp}[3]{Figs.~\ref{#1}({#2}) and ({#3})}
\newcommand{\Figs}[2]{Figs.~\ref{#1} and \ref{#2}}

\newcommand{\Tab}[1]{Table~\ref{#1}}

\newcommand{\aver}[1]{\langle #1\rangle}

\newcommand{\meanrho}{\overline{\rho}}

{}
{}
{}

{}
{}
{}
{}
{}
{}
{}
{}
{}
{}
{}
{}
{}
{}
{}
{}
{}

{}

{}

{}
{}
{}

\newcommand{\kk}{\bm{k}}

\newcommand{\BB}{\bm{B}}

\newcommand{\JJ}{\bm{J}}

\newcommand{\AAA}{\bm{A}}

\newcommand{\uu}{\bm{u}}

\newcommand{\nab}{{\bm{\nabla}}}

\newcommand{\SSSS}{\mbox{\boldmath ${\sf S}$} {}}

\newcommand{\DD}{{\rm D} {}}

\newcommand{\dd}{{\rm d} {}}

\newcommand{\const}{{\rm const}  {}}

\def\la{\mathrel{\mathchoice {\vcenter{\offinterlineskip\halign{\hfil
$\displaystyle##$\hfil\cr<\cr\sim\cr}}}
{\vcenter{\offinterlineskip\halign{\hfil$\textstyle##$\hfil\cr<\cr\sim\cr}}}
{\vcenter{\offinterlineskip\halign{\hfil$\scriptstyle##$\hfil\cr<\cr\sim\cr}}}
{\vcenter{\offinterlineskip\halign{\hfil$\scriptscriptstyle##$\hfil\cr<\cr\sim\cr}}}}}

\def\Sp{\mbox{\rm Sp}}

\def\EEK{{\cal E}_{\rm K}}
\def\EEM{{\cal E}_{\rm M}}

\def\HHM{{\cal H}_{\rm M}}

\def\EK{E_{\rm K}}
\def\EM{E_{\rm M}}

\def\cs{c_{\rm s}}

\def\xiM{\xi_{\rm M}}

\def\vAz{v_{\rm A0}}

\def\kp{k_{\rm p}}
\def\HM{H_{\rm M}}
\def\EM{E_{\rm M}}

\def\kB{k_{\rm B}}

\def\kB{k_{\rm B}}

\def\Brms{B_{\rm rms}}

\def\urms{u_{\rm rms}}

\def\muM{\mu_{\rm M}}

\def\half{{\textstyle{1\over2}}}

\newcommand{\s}{\,{\rm s}}

\newcommand{\cm}{\,{\rm cm}}

\def\apj{Astrophys. J.}

\def\prd{Phys. Rev. D}

\def\prl{Phys. Rev. Lett.}

\def\aap{Astron. Astrophys.}
\def\mnras{Mon. Not. R. Astron. Soc.}

\def\jcap{J. Cosmol. Astropart. Phys.}
\def\jhep{J. High. Energy Phys.}

\def\joss{J. Open Source Softw.}
\definecolor{darkgreen}{RGB}{0, 100, 0}

\begin{document}

\title{Chiral magnetohydrodynamics with zero total chirality}

\date{\today}
\preprint{NORDITA-2023-014, RESCEU-5/23, KEK-TH-2504, MS-TP-23-13}

\author{Axel Brandenburg}
\affiliation{Nordita, KTH Royal Institute of Technology and Stockholm University, 10691 Stockholm, Sweden}
\affiliation{The Oskar Klein Centre, Department of Astronomy, Stockholm University, AlbaNova, SE-10691 Stockholm, Sweden}
\affiliation{School of Natural Sciences and Medicine, Ilia State University, 0194 Tbilisi, Georgia}
\affiliation{McWilliams Center for Cosmology and Department of Physics, Carnegie Mellon University, Pittsburgh, Pennsylvania 15213, USA}

\author{Kohei Kamada}
\affiliation{Research Center for the Early Universe (RESCEU), Graduate School of Science,\\ The University of Tokyo, Hongo 7-3-1, Bunkyo-ku, Tokyo 113-0033, Japan}\

\author{Kyohei Mukaida}
\affiliation{KEK Theory Center, Tsukuba 305-0801, Japan}
\affiliation{Graduate University for Advanced Studies (Sokendai), Tsukuba 305-0801, Japan}

\author{Kai Schmitz}
\affiliation{Institute for Theoretical Physics, University of M\"unster, 48149 M\"unster, Germany}

\author{Jennifer Schober}
\affiliation{Institute of Physics, Laboratory of Astrophysics, \'Ecole Polytechnique F\'ed\'erale de Lausanne (EPFL), 1290 Sauverny, Switzerland}

\begin{abstract}
We study the evolution of magnetic fields coupled with chiral fermion
asymmetry in the framework of chiral magnetohydrodynamics with
zero initial total chirality.
The initial magnetic field has a turbulent spectrum peaking at a certain
characteristic scale and is fully helical with positive helicity.
The initial chiral chemical potential is spatially uniform and negative. 
We consider two opposite cases where the ratio of the length scale of
the chiral plasma instability (CPI) to the characteristic scale of the
turbulence is smaller and larger than unity.
These initial conditions might be realized in cosmological models,
including certain types of axion inflation.
The magnetic field and chiral chemical potential evolve with inverse
cascading in such a way that the magnetic helicity and chirality cancel
each other at all times, provided there is no spin flipping.
The CPI time scale is found to determine mainly the time when the magnetic
helicity spectrum attains negative values at high wave numbers.
The turnover time of the energy-carrying eddies, on the other hand,
determines the time when the peak of the spectrum starts to shift to
smaller wave numbers via an inverse cascade.
The onset of helicity decay is determined by the time when the chiral
magnetic effect becomes efficient at the peak of the initial magnetic
energy spectrum, provided the CPI does not grow much.
When spin flipping is important, the chiral chemical potential vanishes
at late times and the magnetic helicity becomes constant, 
which leads to a faster increase of the correlation length.
This is in agreement with what is expected from magnetic helicity
conservation and also happens when the initial total chirality is
imbalanced.
Our findings have important implications for baryogenesis after axion inflation.
\end{abstract}

\maketitle

\section{Introduction}
\label{Intro}

Relativistic plasmas are described by the evolution equations of
chiral magnetohydrodynamics (MHD) \cite{Son:2009tf,Neiman:2010zi,BFR12, BFR15, Pavlovic+17,
Roga_etal17, Bran_etal17, Schober+18, DelZanna+Bucciantini18}.
Chirality enters in two distinct ways: first, through a nonvanishing
chiral chemical potential, $\tilde{\mu}_5$, and second, through nonvanishing
magnetic helicity density, $\AAA\cdot\BB$, where $\BB=\nab\times\AAA$
is the magnetic field expressed in terms of the vector potential $\AAA$. 

It has been known for some time that fermion chirality can be transferred
into magnetic helicity and vice versa through the chiral anomaly~\cite{Adler:1969gk,Bell:1969ts}.
The transfer of fermion chirality to magnetic helicity occurs through
an instability \cite{JS97} known as the chiral plasma instability (CPI)
\cite{Akamatsu:2013pjd}.
This instability is the fastest at a specific wave number, whose value
depends on the chiral chemical potential.
The transfer from magnetic helicity to chiral chemical potential does
not involve any instability, but occurs just through a nonvanishing
nonlinear source term in the evolution equation for the chiral chemical
potential \cite{BFR12,Hirono+15, Schober+2020}.
These differences in the evolutions of the chiral chemical potential
and magnetic field can lead to nontrivial dynamics, which has triggered
a lot of research \cite{Manuel+15, Domcke+19b, Domcke+20}.
Since fermion chirality is tightly related to the baryon and lepton asymmetries
at high temperature in the early Universe,
their co-evolution with magnetic helicity in the context of cosmology
has also been extensively studied \cite{Giovannini:1997eg, Giovannini:1997gp,
Dvornikov:2011ey, Tashiro:2012mf, Long:2013tha, Fujita:2016igl,
Kamada:2016eeb, Kamada:2016cnb, Pavlovic:2016gac}.

Previous investigations mostly assumed an initial imbalance between
fermion chirality and magnetic helicity.
In many investigations, the initial fermion chirality is nonvanishing
while initial magnetic helicity is zero or vice versa.
This can lead to a conversion of fermion chirality to a maximally
helical magnetic field \cite{BFR12}.
Also just spatial fluctuations can lead to magnetic field production
\cite{Schober+2022a,Schober+2022b}.
Such chiral asymmetry, which can trigger the CPI, could be
generated \cite{GarciaBellido+99, Kamada:2018tcs, Domcke:2020quw}
in GUT baryogenesis in the early Universe~\cite{Yoshimura:1978ex,
Dimopoulos:1978kv, Toussaint:1978br, Weinberg:1979bt, Barr:1979ye}
or weak interactions in compact stars \cite{Ohnishi:2014uea,
Grabowska+15, Masada+18, Dvornikov+20, Matsumoto:2022lyb} (see also
Ref.~\cite{Kamada:2022nyt} and references therein).
However, numerical studies on other interesting initial conditions are
still lacking, where fermion chirality is exactly opposite to magnetic
helicity.
Such an initial condition is expected if the chiral symmetry in the fermion sector is only
broken through the topological density, $\partial_\mu J^\mu_5 = -
e^2 F_{\mu\nu} \tilde F^{\mu\nu} / (8 \pi^2 \hbar^2 c)$,
or the chiral anomaly~\cite{Adler:1969gk,Bell:1969ts}, 
with $J_5^\mu$ being
the chiral current and $e^2 F_{\mu\nu} \tilde F^{\mu\nu} / (8 \pi^2 \hbar^2 c)$
being the topological density.
Since the topological density can be written as a total derivative of
the magnetic helicity density, the sum of chiral asymmetry and magnetic
helicity vanishes when they are generated \cite{domcke2018}.

Configurations with vanishing total chirality are interesting not only
in the context of chiral MHD, but also in particle physics and cosmology.
At high enough temperatures, realized in the early Universe, the
electron Yukawa interaction becomes inefficient for
$T \gtrsim 10^5\, \mathrm{GeV}$ \cite{Campbell:1992jd,Bodeker:2019ajh}.
There we find the conservation of total chirality because of 
$\partial_\mu J^\mu_{e_R} = - g_Y^2 Y_{\mu\nu} \tilde Y^{\mu\nu} / (16 \pi^2 \hbar^2 c)$ 
with $J^\mu_{e_R}$ being the right-handed electron current and
$Y^{\mu\nu}$ being the field strength of the hypercharge gauge field
with gauge coupling $g_Y$.
For instance, in a certain class of axion inflation, configurations with
zero net chirality are generated during inflation~\cite{domcke2018},
which can be the origin of the observed baryon asymmetry of the
Universe \cite{Jimenez:2017cdr,Domcke+19, Domcke:2022kfs} and it
could explain the proposed intergalactic magnetic field; see, however,
Ref.~\cite{Kamada:2016cnb} for the baryon overproduction problem and
Ref.~\cite{Kamada:2020bmb} for the too large baryon isocurvature problem.
The main purpose of this paper is to perform a full numerical chiral MHD
simulation under the initial condition of vanishing total chirality and
provide a better understanding of the nonlinear dynamics in this case.

Before we begin our investigations, it is useful to recall the main
findings of earlier work where the total chirality was mostly different from zero.
Following the work of Ref.~\cite{Hirono+15}, who studied a system
consisting of the gauge field and the chiral chemical potential,
but without fluid velocity fields, and with the initial condition
$\aver{\AAA\cdot\BB} \not =0, \tilde{\mu}_5=0$, three stages can be identified:
(i) exponential decline of the magnetic helicity together with
an increase of $\tilde{\mu}_5$, followed by
(ii) a continued decrease of the typical peak wave number $\kp$,
while $\tilde{\mu}_5$ stays at its maximum value with $\aver{\AAA\cdot\BB}$
being essentially zero, and (iii) a phase when all the fermion chirality
$\tilde{\mu}_5$ gets transferred back to magnetic helicity.
As expected, owing to magnetic helicity conservation, and because the
magnetic field from the CPI is maximally helical,
the magnetic energy density $\aver{\BB^2}/2$ decays at late times such
that $\aver{\BB^2}\xiM\approx\const$, where $\xiM\equiv\kp^{-1}$ is the
magnetic correlation length.
In other words, both $\aver{\BB^2}$ and $\kp$ decay in the same fashion,
but, unlike the expected $t^{-2/3}$ scaling found previously for helical
turbulence \citep{Hat84,BM99,Kahn_etal13,BK17}, the authors
of Ref.~\cite{Hirono+15} find a $t^{-1/2}$
scaling both for $\aver{\BB^2}$ and $\kp$.
For sufficiently strong initial magnetic fields, the magnetic Reynolds
number can be much larger than unity and the eddy turnover scale much
longer than the estimated inverse peak momentum scale, if equipartition
between the magnetic field and fluid velocity is established.
This suggests that the effect of the fluid velocity cannot be negligible
in general.

The earlier analytic study of Ref.~\cite{Hirono+15} was revisited using
direct numerical simulations of chiral MHD \cite{Schober+2020}.
At large magnetic Reynolds numbers, the authors found clear evidence
for a $t^{-2/3}$ scaling of both $\aver{\BB^2}$ and $\kp$ at late times.
They also found that the initial evolution is not exponential, as
suggested in Ref.~\cite{Hirono+15}, but linear in time.
However, they only considered the case where the initial fermion
chirality was zero.
When it is finite and balancing exactly the magnetic helicity, the
magnetic field decays in a way similar to the case of a strong, nonhelical
field~\cite{BKS23}, where the decay is governed by the conservation of the Hosking
integral \cite{Hosking+Schekochihin21, Hosking+Schekochihin22, Zhou+22}.
This integral describes the strength of magnetic helicity fluctuations
on different length scales and has the dimensions of $\cm^9\s^{-4}$,
which implies the scalings $\xiM\propto t^{4/9}$ and
$\aver{\BB^2}\propto t^{-10/9}$ \cite{Hosking+Schekochihin21, Zhou+22, BL23}.
The general validity of the Hosking integral was further demonstrated by
applying a corresponding analysis to the decay of a nonhelical magnetic
field in neutron star crusts \cite{Bra23}, where the magnetic field
evolution is covered by the Hall effect \cite{GR92}.

Our goal here is to bridge the gap between the two extremes, where
the initial chirality is either only in the fermions or only in the
magnetic field, and to consider the intermediate case
where fermion chirality and magnetic helicity balance to zero, 
extending the study of the present authors \cite{BKS23}.
This is another case where the decay of $\aver{\BB^2}$ and $\kp$ are
described by a correspondingly adapted Hosking integral of the total
chirality.
In the following, we therefore refer to the Hosking integral with the
chiral chemical potential included as the ``adapted'' Hosking integral;
see Ref.~\cite{BKS23} for detail.

As mentioned above, our findings on the evolution of the system with
vanishing total chirality has significant impact on the present baryon
asymmetry of the Universe.
Another goal of the present paper is to clarify how the non-trivial
co-evolution of the magnetic field and fermion chirality affect the
model space of axion inflation consistent with the present Universe,
which has not been explored before.

We begin, by presenting the basic equations and the mathematical setup
of our simulations in \Sec{ChiralMHD}.
We then discuss the parameter dependence of characteristic time scales,
consider the effect of spin flipping, and finally cases where the
perfectly vanishing chirality balance is relaxed in \Sec{Results}.
Applications to the early Universe are discussed in \Sec{Application}.
Conclusions are presented in \Sec{Conclusions}.

\section{Chiral magnetohydrodynamics}
\label{ChiralMHD}

\subsection{Chiral magnetic effect}

Using Lorentz-Heaviside units, the Amp\`{e}re-Maxwell equation 
for the QED-like model in the MHD limit (omitting the displacement current) reads
\begin{equation}
{\bm \nabla} \times {\bm B} = \frac{1}{c} {\bm J}. 
\end{equation}
The electric current ${\bm J}$ is the sum of the Ohmic current and 
the chiral magnetic effect (CME) \cite{Vilenkin80, Alekseev:1998ds, Fukushima:2008xe},
\begin{equation}
{\bm J} = \frac{\sigma}{c} \left(c{\bm E} + {\bm u} \times {\bm B}\right) + \frac{e^2}{2 \pi^2 \hbar^2 c} \tilde{\mu}_5 {\bm B}, 
\label{current}
\end{equation}
where $\sigma$ is the electric conductivity (or inverse magnetic diffusivity) and
we consider the case with $\tilde{\mu}_5 \ll (e^2/\hbar c) \kB T$.
By rewriting $c{\bm E} = - \partial {\bm A}/\partial t$ in the Weyl gauge,
$e^2/4\pi \hbar c\equiv \alpha$, \Eq{current} is rewritten as
\begin{equation}
\frac{\partial {\bm A}}{\partial t} = \frac{c^2}{\sigma} \left( \mu_5 {\bm B} - \nab\times{\bm B}\right) + {\bm u} \times {\bm B} ,
\label{dAdt}
\end{equation} 
where we defined \cite{Roga_etal17}
\begin{equation}
\mu_5 \equiv \frac{2\alpha}{\pi\hbar c}\tilde{\mu}_5.
\label{mu5-tilde}
\end{equation}
This expression agrees with Eq.~(32) of Ref.~\cite{Roga_etal17},
except for a factor of 2 resulting from a corresponding
1/2 factor in our adopted definition
$\tilde{\mu}_5=(\tilde{\mu}_{\rm R}-\tilde{\mu}_{\rm L})/2$ in
terms of the chemical potentials for right- and left-handed fermions
\cite{Kamada:2022nyt}.
The additional $4\pi$ factor in the numerator of the expression in
Ref.~\cite{Roga_etal17} is a consequence of their use of cgs units.

\subsection{Model description and basic equations}
\label{ModelDescription}

We perform simulations in a cubic domain of size $L^3$ with side lengths
$L$ and triply-periodic boundary conditions.
The mass in the domain is therefore constant, so the mean density
$\meanrho$ is always equal to its initial value $\rho_0$
and put to unity in all cases.
The lowest wave number in the domain is $k_1=2\pi/L$.
Using $N^3$ mesh points, the largest wave number in the simulations is
the Nyquist wave number $k_{\rm Ny}=k_1 N/2$.

In the following, we set $c=1$, so $\JJ=\nab\times\BB$.
To include the effects of the cosmic expansion with scale factor
$a(t)\propto t^{1/2}$ in the radiation-dominated era, which we assume to
be a spatially flat Friedmann Universe, we use correspondingly scaled
quantities and conformal time, $\eta(t)=\int\dd t/a(t)$, in which the
evolution equations of MHD are the same as in the absence of expansion
\cite{BEO96}.
In order to obtain the physical quantities, we can simply normalize the
corresponding comoving quantities with the appropriate powers of the
scale factor $a$.
Furthermore, using $\lambda=3 \hbar (2 \alpha/ \pi \kB T)^2$ and including
spin flipping and spatial diffusion, our chiral anomaly equation is
\begin{equation}
\frac{\partial\mu_5}{\partial\eta} + {\bm \nabla}\cdot(\mu_5 {\bm u}) = \frac{\lambda}{\sigma}
\left({\bm J}-\mu_5 {\bm B} \right) \cdot {\bm B}+D_5\nabla^2\mu_5-\Gamma\mu_5,
\label{dmudt}
\end{equation}
where $D_5$ is an empirical diffusion coefficients for the chiral
chemical potential.
Here we used the relationship 
between the chiral chemical potential and the number density, 
\begin{equation}
n_5 \equiv n_R-n_L =  2 \times \frac{\tilde{\mu}_5}{6\hbar ^3} (\kB T)^2
=\frac{\pi \mu_5}{6 \alpha \hbar^2} (\kB T)^2,
\end{equation}
and used $J_5^\mu=(n_5,n_5 {\bm u}-D_5\nab n_5)$ for the chiral 4-current.

Owing to the chiral anomaly \citep{Adler:1969gk,Bell:1969ts},
the total chirality is conserved in the absence of spin flipping
interaction \citep{BFR12, Roga_etal17}.
It is then convenient to introduce the mean magnetic chirality
equivalent as
\begin{equation}
\aver{\mu_{\rm M}}\equiv\half\lambda\aver{\AAA\cdot\BB}, 
\label{muMdef}
\end{equation}
so that the conservation law derived from \Eqs{dAdt}{dmudt} can be
stated in the form
\begin{equation}
\mu_{\rm tot}=\aver{\mu_5}+\aver{\mu_{\rm M}}=\const.
\label{eq:mutot}
\end{equation}

We complement \Eqs{dAdt}{dmudt} by the momentum and continuity equations
\citep{Roga_etal17, Bran_etal17, BHKRS21}
\begin{align}
\frac{\DD\uu}{\DD\eta}&=\frac{2}{\rho}\nab\cdot\left(\rho\nu\SSSS\right)
-\frac{1}{4}\nab\ln\rho+\frac{\uu}{3}\left(\nab\cdot\uu
+\uu\cdot\nab\ln\rho\right)
\nonumber \\
&-\frac{\uu}{\rho}\left[\uu\cdot(\JJ\times\BB)+\eta \JJ^2\right]
+\frac{3}{4\rho}\JJ\times\BB,
\label{dudt} \\
\frac{\partial\ln\rho}{\partial\eta}
&=-\frac{4}{3}\left[\nab+ \left(\nab\ln\rho\right)\right] \cdot \uu
+\frac{1}{\rho}\left[\uu\cdot(\JJ\times\BB)+\eta \JJ^2\right]\!,
\nonumber
\end{align}
where $\DD/\DD\eta\equiv\partial/\partial\eta+\uu\cdot\nab$ is the
advective derivative, ${\sf S}_{ij}=(\partial_i u_j+\partial_j u_i)/2
-\delta_{ij}\nab\cdot\uu/3$ are the components of the rate-of-strain
tensor, $\nu$ is the viscosity, and $p$ is the
pressure, which is assumed to be proportional to the density, i.e.,
$p=\rho\cs^2$, with $\cs = 1/\sqrt{3}$ being the sound speed
for the ultrarelativistic fluid.
The ratio of viscosity to magnetic diffusivity, $\nu\sigma$, is the
magnetic Prandtl number, which we choose to be unity in all cases.
Likewise, we choose the ratio $\nu/D_5$ to be unity in all cases.

For all our simulations, we use the {\sc Pencil Code} \cite{JOSS},
where the relevant equations are readily implemented.
We use $N^3=1024^3$ mesh points for most of the runs, and $N^3=2048^3$
mesh points for one particular run.
In a small number of cases, we have included the slope-limited diffusion
(SLD) scheme of Ref.~\cite{Rem14,NewsletterSLD}.
In those cases, SLD acts in addition to the ordinary viscous and diffusive
processes stated in the equations above, but prevents the code from
crashing during an early more violent phase when the mesh resolution is
insufficient to dissipate the energy at high wave numbers.
At later times, however, this additional numerical device has little
effect.
Below, we demonstrate in one case that the solutions with and without
SLD yield the same result.

\subsection{Diagnostic quantities}
\label{DiagnosticQuantities}

We introduce two characteristic times in our simulations, which are the
time scale of the CPI and the magnetic diffusion time,
\begin{equation}
\eta_{\rm CPI}=\sigma\mu_{50}^{-2}\quad\mbox{and}\quad
\eta_{\rm diff}=\sigma k_0^{-2},
\label{etaCPIdiff}
\end{equation}
respectively.
Here, $k_0$ is the initial value of the peak wave number $\kp$.
The ratio $(\eta_{\rm diff}/\eta_{\rm CPI})^{1/2}=|\mu_{50}|/k_0$
characterizes the degree of scale separation between the scales of
magnetic helicity and fermion chirality.
We also define the turnover time of the energy-carrying eddies,
which would determine the onset of turbulent inverse cascading,
\begin{equation}
\eta_{\rm turb}=\left(\urms^{\max} k_0\right)^{-1}, 
\label{etaturb}
\end{equation}
where $\urms^{\max}$ is the maximum value (in time) of the rms velocity.

Next, we introduce several parameters with a dimension of velocity.
The nature of the CPI is characterized by the following parameters
\cite{Bran_etal17}
\begin{equation}
v_\lambda=|\mu_{50}|/(\meanrho\lambda)^{1/2}\quad\mbox{and}\quad
v_\mu=|\mu_{50}|/\sigma. 
\end{equation}
The former represents the ratio of the length scale of the magnetic field
at saturation of the CPI to the CPI time scale,
while the latter represents the ratio of the length scale of the initial instability
to the CPI time scale.
The ratio $v_\lambda/v_\mu=\sigma/(\meanrho\lambda)^{1/2}$
characterizes the length of the $k^{-2}$ spectrum that develops if the
CPI operates without a strong pre-existing field \cite{Bran_etal17}.
In the unbalanced case, $\urms^{\max}$ is approximated by $v_\lambda$.
In the present case, however, it does not seem to play any role.
Instead, to compute $\urms^{\max}$, we approximate the velocity field by
the initial magnetic field such that $\Brms^2 \simeq {\bar \rho} \urms^2$.
Using \Eqs{muMdef}{eq:mutot}, we estimate
\begin{equation}
    \Brms^{(0)} \approx(k_0|\aver{\AAA\cdot\BB}|)^{1/2} \approx \left(\frac{2k_0|\mu_{50}|}{\lambda }\right)^{1/2}, 
\end{equation}
which thus defines a new quantity $\tilde{v}_\lambda$ as
\begin{equation} \label{tildevlam}
\tilde{v}_\lambda \equiv \left(\frac{2k_0|\mu_{50}|}{{\bar \rho} \lambda} \right)^{1/2}
\quad \left(\approx \frac{\Brms^{(0)} }{{\bar \rho}^{1/2}} \right). 
\end{equation}
A predictive estimate for the turnover time of the energy-carrying eddies
is thus
\begin{equation}
\eta_\lambda=(\tilde{v}_\lambda k_0)^{-1} = \left(\frac{{\bar \rho} \lambda}{2 k_0^3 |\mu_{50}|}\right)^{1/2},
\label{etalam}
\end{equation}
which is later used to predict the time when the inverse cascade sets in.

In this work, an important diagnostics is the magnetic energy spectrum, $\EM(k)$.
It is normalized such that $\int\EM(k)\,\dd k=\aver{\BB^2}/2\equiv\EEM$ 
where $\EEM$ is the magnetic energy density\footnote{
In terms of the mode function in the polarization basis,
${\bm A}({\bm x},t) \equiv \int d^3 k/(2\pi)^{3/2} \sum_{\lambda=\pm}
A_\lambda({\bm k},t) {\bm e}^\lambda ({\bm k}) e^{i {\bm k} {\bm x}}$,
$\EM$ is given as $\EM(k)= \sum_{s=\pm} (k^4/4\pi^2) |A_s(k)|^2$. 
We also have $\HM^s (k) = (k^3/2 \pi^2) |A_s(k)|^2$ and
$\HM (k) = (k^3/2 \pi^2) \sum_{s=\pm} s|A_s(k)|^2$.}.
The kinetic energy spectrum $\EK(k)$ is defined similarly, i.e.,
$\int\EK(k)\,\dd k=\aver{\rho\uu^2}/2\equiv\EEK$.
We also define the magnetic helicity spectrum $\HM(k)$, which is
normalized such that $\int \HM(k) \, \dd k = \aver{\AAA\cdot\BB}$.
In our simulations, $k|\HM(k)|/2$ approaches $\EM(k)$ near the maximum.
In fact, the spectra $\HM(k)$ and $\EM(k)$ satisfy the realizability
condition \cite{Mof78},
\EQ
k|\HM(k)|/2\le \EM(k).
\label{realizability}
\EN
When this inequality is saturated for specific wave numbers, we say that
the magnetic field is locally fully helical.

After some time, the magnetic helicity spectrum is characterized by two
subranges, one with positive and one with negative values of $\HM(k)$,
which are separated by the wave number $k_\pm(\eta)$, where the sign
changes.
In addition to the evolution of $k_\pm(\eta)$, we characterize the
spectrum and its evolution by the numbers  $k_{\rm I}(\eta)$ and
$k_{\rm II}(\eta)$, which are the wave numbers of the first positive
and second negative peak of $\HM(k)$.\footnote{In the present work,
$k_{\rm I}\approx\kp\equiv\xiM^{-1}$, but the latter is based on the
magnetic energy spectrum while the former is based on the magnetic
helicity spectrum.}
The intermediate wave number $k_\pm(\eta)$ is numerically often better determined
than $k_{\rm II}(\eta)$, especially at early times.

The wave number of the first peak of the spectrum is close to the
initial inverse correlation length,
\EQ
\xiM=\EEM^{-1}\int k^{-1}\EM(k)\,\dd k.
\EN
In fully helical turbulence, the value of $\xiM(\eta)$ tends to increase
with time in a power law fashion, $\xiM\propto\eta^q$, where $q=4/9$ in
our cases of balanced chirality \cite{BKS23}; see also \Sec{IIE}.
Note that in our setup the positive helicity modes always dominate
the energy density of the magnetic field, and hence approximately we have
$\xiM \simeq k_{\rm I}^{-1}$.

It is convenient to introduce the mean magnetic chirality for the positive
helicity modes for $k < k_\pm$ and the negative ones for $k>k_\pm$ as
\begin{align}
\label{muMp} \aver{\mu_{\rm M}^+} &= \frac{\lambda}{2} \int_0^{k_\pm} H_{\rm M}(k)\,\dd k, \\
\label{muMm} \aver{\mu_{\rm M}^-} &= -\frac{\lambda}{2} \int_{k_\pm}^\infty  H_{\rm M}(k)\,\dd k.
\end{align}
The conservation law takes then the form
\begin{equation}
\aver{\mu_5}+\aver{\mu_{\rm M}^+}-\aver{\mu_{\rm M}^-}=\mu_{\rm tot},
\end{equation}
where $\mu_{\rm tot}=\mu_{50}+\mu_{\rm M0}=\mu_{50}+\mu_{\rm M0}^+
-\mu_{\rm M0}^-$ is given by the initial values.

When we study the effect of spin flipping, we invoke a nonvanishing flipping
rate with
\EQ
\Gamma=
\left\{
\begin{aligned}
\Gamma_{\rm f0} & \quad\mbox{for}\quad 
\eta_{\rm flip}\leq\eta\leq\eta_{\rm off} \\
0\;\; & \quad\mbox{otherwise},\\
\end{aligned}
\right.
\label{etaflip}
\EN
where $\eta_{\rm flip}$ denotes the time when spin flipping is turned
on, and in a few cases we allow for a finite value of $\eta_{\rm off}$,
which denotes the time when spin flipping is later turned off again.
In the context of early Universe cosmology, we have indeed a constant flipping rate
for the thermal Yukawa interaction~\cite{JS97}. 
A sudden turnoff is introduced for illustration, 
but such a transition can be realized by introducing new physics. 
Going into technical detail is beyond the scope of this paper. 

\subsection{Initial conditions}
\label{InitialConditions}

In our numerical experiments, the initial magnetic field is fully helical
with positive magnetic helicity and random phases.
The initial magnetic energy spectrum is a broken power law
\begin{equation}
\EM(k,\eta_0)\propto \begin{cases}
k^4 & \mbox{for}\;k<k_0,\\
k^{-5/3} & \mbox{for}\;k>k_0, 
\end{cases}
\label{k0def}
\end{equation}
where the initial peak is identified as $k_0 = k_{\rm I}(\eta_0)$
and $\eta_0$ is the initial time.
The IR spectrum is motivated by causality constraints \cite{DC03},
while the UV spectrum is taken as a Kolmogorov-type spectrum.
The strength of the magnetic field is adjusted such that the initial
magnetic chirality obeys 
$\mu_{\rm M0}=-\mu_{50}$ such that $\mu_\mathrm{tot} =0$.
The chiral chemical potential is initially assumed to be uniformly
distributed in space.
Its initial value is always negative, i.e., $\mu_{50}<0$.
However, even for an initially uniform chiral chemical potential, there
is a specific length scale associated with the value of $\mu_5$ through
the wave number of the most unstable mode of the CPI, $k=|\mu_{50}|/2$. 
The initial velocity is assumed vanishing in all cases.

\subsection{Theoretical predictions} \label{IIE}

As was recently shown in Ref.~\cite{BKS23}, the present case of zero
total chirality, where the magnetic helicity is canceled by fermion
chirality, is remarkably similar to the case of ordinary MHD without
chemical potential and zero magnetic helicity.
In both cases, as already alluded to in the introduction,
one can define a correlation integral of the total chirality,
which is a quantity with dimensions $\cm^9\s^{-4}$
and is dubbed the adapted Hosking integral. 
The evolution of the system can be explained by the conservation 
of this quantity.
With a self-similar evolution of the magnetic spectrum being assumed,
this yields the scalings $\xiM\propto\eta^{4/9}$ and
$\aver{\BB^2}\propto\eta^{-10/9}$ for the typical length scale and the
magnetic energy density, respectively \cite{Hosking+Schekochihin21}.
Note that the conservation of the adapted Hosking integral suggests 
\begin{equation}
\xi_M^5 \langle \BB^2 \rangle^2 = \const, \quad\text{or} \quad k_{\rm I}^{-3} E_M(k_{\rm I})^2 = \const, \label{hoscons}
\end{equation}
if the magnetic energy density is 
dominated by the positive helicity mode, which is peaked at $k=k_{\rm I}$.
For the magnetic field with an IR spectrum $\propto k^4$, as motivated
from the causality constraints, the evolution of the magnetic field
exhibits inverse cascading.

The big difference between ordinary MHD without helicity on the one hand and chiral
MHD with helicity balanced by fermion chirality on the other hand is that in the latter,
both the magnetic helicity and the fermion chirality are decaying, 
which we shall call {\em anomalous chirality cancellation} (ACC).
In the former, by contrast, the Hosking integral based just on the
ordinary magnetic helicity density is conserved.
In the latter, contrary to the naively expected exponential decay of
fermion chirality due to the CPI in chiral MHD, we actually have a much
slower power-law decay proportional to $\eta^{-2/3}$, since the magnetic
helicity is roughly estimated by $\xi_M \langle \BB^2 \rangle$, 
and likewise for $|\aver{\mu_5}|$ \cite{BKS23}. 
Here we have considered the case where
the real space realizability condition
of magnetic helicity \citep{Kahn_etal13}, $|\HHM|\leq2\EEM\xiM$,
is nearly saturated at ACC onset.
Once this power law decay of the chirality starts,  
the CPI rate, $\aver{\mu_5}^2/\sigma$, decays faster than $\eta^{-1}$, 
which suggests that the CPI does not grow anymore. 
Hence, the magnetic energy is always dominated by helicity modes of
the same sign as the initial ones, which, in our case, are positive
helicity modes.

The adapted Hosking integral makes sense only when the communication between 
the helicity and chirality through the CME becomes effective at the characteristic scale.
Therefore we expect that the scaling evolution discussed above starts 
at the time scale of the CME at the peak scale. 
With the evolution equation for the magnetic field, equivalent to Eq.~\eqref{dAdt}, 
\begin{equation}
\frac{\partial {\bm B}}{\partial \eta} = \frac{1}{\sigma} \left[\nabla^2 {\bm B}
+ {\bm \nabla} \times (\mu_5 {\bm B})\right]
+ \nab \times ({\bm u} \times{\bm B}) \label{BEOM}
\end{equation}
(where the second term in the right-hand side represents the CME), 
we estimate $\eta_\mathrm{ACC}$ as the time when the following condition is satisfied:
\begin{equation}
\eta_\mathrm{ACC} \simeq \frac{\sigma}{\mu_5(\eta_\mathrm{ACC}) k_{\rm I}(\eta_\mathrm{ACC})}.
\label{etaacc}
\end{equation}
Note that from Eq.~\eqref{BEOM} we can also confirm that 
the magnetic field has an instability (the CPI) for one of the two
circular polarization modes with $k=|\mu_{50}|/2$ being the
most unstable mode.
The instability rate is roughly given by $\mu_{50}^2/\sigma$, 
which determines $\eta_\mathrm{CPI}$.

\begin{table*}[t]\caption{
Relevant time scales defined in this paper.
}\vspace{12pt}\centerline{\begin{tabular}{llll}
Time scale & Expression & Equation & Explanation \\
\hline
$\eta_{\rm CPI}$  & $\sigma\mu_{50}^{-2}$     & \Eq{etaCPIdiff} & time scale of the CPI \\
$\eta_{\rm diff}$ & $\sigma k_0^{-2}$         & \Eq{etaCPIdiff} & magnetic diffusion time \\
$\eta_{\rm turb}$ & $(\urms^{\max} k_0)^{-1}$ & \Eq{etaturb}    & turnover time of the energy-carrying eddies \\
$\eta_\lambda$    & $(\tilde{v}_\lambda k_0)^{-1} = [{\bar \rho} \lambda/(2 k_0^3 |\mu_{50}|)]^{1/2}$ & \Eq{etalam} & {\em predicted} turnover time of the energy-carrying eddies \\
$\eta_\mathrm{ACC}  $ & $\sigma/[\mu_5(\eta_\mathrm{ACC}) k_{\rm I}(\eta_\mathrm{ACC})]$ & \Eq{etaacc} & onset time of the ACC \\
\label{TAnalyticTimeScale}\end{tabular}}\end{table*}

The evolution of the system is classified into two cases, 
determined by the comparison between $\eta_\lambda$ and $\eta_\mathrm{ACC}$
estimated by the initial conditions of $k_{\rm I}$ and $\mu_5$. 
For relativistic plasmas with ${\bar \rho}\simeq (\pi^2 g_*/30)T^4$, where $g_*$ is number
of the relativistic degrees of freedom and $\sigma \simeq T/(\alpha \log \alpha^{-1})$
\cite{Baym:1997gq,Arnold:2000dr},
we have $\eta_\mathrm{ACC} < \eta_\lambda$ for $k_0 \ll |\mu_{50}|$ 
[more precisely, $k_0 \ll ({\bar \rho} \lambda/ 4 \sigma^2) |\mu_{50}|$,
which is independent of temperature], and vice versa.
For $k_0 \ll |\mu_{50}|$, we have the following estimates
for the evolution of the system:
\begin{enumerate}
    \item The system is frozen when $\eta<\eta_\mathrm{CPI}$. 
    \item The CPI starts to grow at $\eta \simeq \eta_\mathrm{CPI}$. 
    \item If the CPI does not sufficiently amplify the negative helicity modes
    such that $k_{\rm I}$ is unchanged, the chiral chemical potential starts to decay at $\eta=\eta_\mathrm{ACC} (>\eta_\mathrm{CPI})$ with
    \begin{equation} \label{etaACCcase1}
        \eta_\mathrm{ACC} \simeq \frac{\sigma}{|\mu_{50}| k_0}
    \end{equation}
    in a mild way.
    \item When $\eta \simeq \eta_\lambda (>\eta_\mathrm{ACC})$, the system starts to evolve
    according to the scaling law found in Ref.~\cite{BKS23}, 
    \begin{align}
        &k_{\rm I} \propto \eta^{-4/9}, \quad \EEM \propto \eta^{-10/9},\quad\mbox{and}\\
        &\aver{\mu_5}=-\aver{\mu_{\rm M}^+}+\aver{\mu_{\rm M}^-} \propto \eta^{-2/3}. 
    \end{align}
\end{enumerate}
In the case of mild hierarchy\footnote{With ``mild hierarchy'', 
we have in mind a modest scale separation $|\mu_{50}|/k_0 = {\cal O}(10)$.}, $|\mu_{50}| \gtrsim k_0$, 
the case we mainly study in the next section,  
the assumption of inefficient CPI is guaranteed.

%FIG1
\begin{figure*}[t!]\begin{center}
\includegraphics[width=\textwidth]{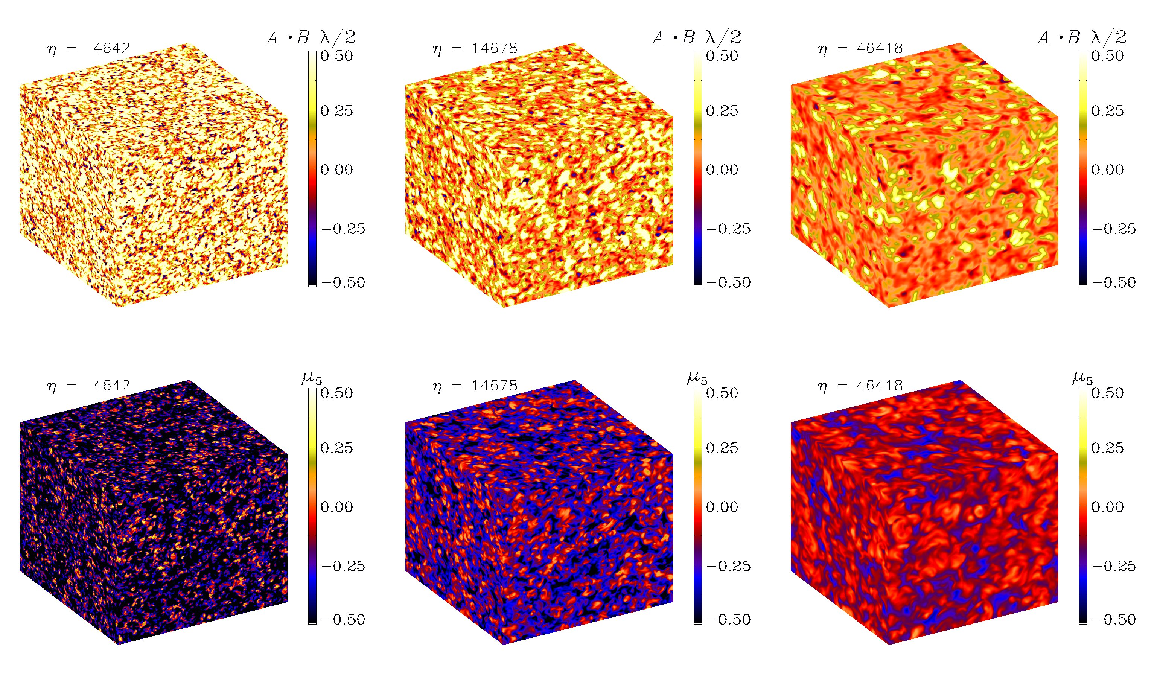}
\end{center}\caption[]{
Visualizations of $\AAA\cdot\BB\,\lambda/2$ (upper row) and $\mu_5$
(lower row) on the periphery of the computational domain for Run~O at
$\eta\approx4600$ (left) 15,000 (middle), and 46,000 (right).
}\label{AB}\end{figure*}

For $k_0 \gg|\mu_{50}|$, on the other hand, we expect the following evolution of the system. 
\begin{enumerate}
    \item The system is frozen at $\eta<\eta_\lambda$. 
    \item The magnetic field evolves according to the inverse cascade
    at $\eta \simeq \eta_\lambda$
    in a similar way as the usual inverse cascade for a nonchiral helical 
    magnetic field, 
    \begin{align}
        &k_{\rm I} \propto \eta^{-2/3}, \quad \EEM \propto \eta^{-2/3},\quad\mbox{and}\\
        &\aver{\mu_5}=-\aver{\mu_{\rm M}^+}+\aver{\mu_{\rm M}^-} = \mathrm{const}, 
    \end{align}
    since the CME is not effective at $k \simeq k_{\rm I}$ so that
    the magnetic helicity and chirality are individually conserved. 
    Since $k_{\rm I}$ decays, eventually it becomes smaller than $|\mu_5|$,
    and the system enters the regime similar to the previous case.
    \item 
    If the CPI does not sufficiently amplify the negative helicity modes, the CME becomes effective at $\eta \simeq \eta_\mathrm{ACC}$ $(>\eta_\lambda)$, 
    which is now evaluated as
    \begin{equation}
    \eta_\mathrm{ACC} \equiv \frac{\sigma^3}{|\mu_{50}|^3 k_0^3} \eta_\lambda^{-2}
    \simeq \frac{2\sigma^3}{{\bar \rho} \lambda \mu_{50}^2 }.
    \label{etaacck0>mu50}
    \end{equation}   
    Here, we have used \Eq{etaacc} and
    $k_{\rm I}(\eta) = k_0 (\eta/\eta_\lambda)^{-2/3}$,
    as well as \Eq{tildevlam}.
    When $\eta>\eta_\mathrm{ACC}$ we have the inverse cascade with the conservation of the adapted Hosking integral, 
    \begin{align}
        & k_{\rm I} \propto \eta^{-4/9}, \quad \EEM \propto \eta^{-10/9},\quad\mbox{and} \label{ACCscaling1}\\
        & \aver{\mu_5}=-\aver{\mu_{\rm M}^+}+\aver{\mu_{\rm M}^-} \propto \eta^{-2/3}. \label{ACCscaling2}
    \end{align}
    Note that in this case, the CPI would not grow much due to the earlier
    onset of the chirality decay. 
\end{enumerate}
The assumption of inefficient CPI is guaranteed if the mild hierarchy
$|\mu_5| \gtrsim k_\text{I}$ at $\eta_\mathrm{CPI}$ or $\eta_\mathrm{ACC}$ still holds. 
However, we have $|\mu_5|/k_\text{I}|_{\eta_\mathrm{CPI}} = (2 \sigma^2/{\bar \rho} \lambda)^{1/3}$,
which is ${\cal O}(10^2)$ for the plasma of the Standard Model (SM)  of particle physics
with $\alpha \simeq 10^{-2}$ and $g_*\simeq 10^2$. 
In this case, there is a relatively large hierarchy between the 
chiral chemical potential and the peak scale of the magnetic field. 
We then might expect an earlier onset of the chirality decay
triggered by the CPI. 
Namely, we have another possibility of the scaling law, which is a
modification of step~3 discussed above, so we refer to
it as step~$3^\prime$, i.e.,
\begin{enumerate}
\renewcommand{\labelenumi}{{\arabic{enumi}$.\!^\prime$}}
\setcounter{enumi}{2}
    \item After some epoch of the onset of CPI, 
    \begin{equation}
        \eta\simeq \eta_{5\mathrm{dec}} = c_A \,\eta_\mathrm{CPI}, \quad \text{with} \quad c_A = \mathcal{O} (10),  \label{eta5dec}
    \end{equation} the system enters the regime of the ACC. 
    If the conservation of the adapted Hosking integral, 
    with the rapid communication between the chirality and helicity 
    through the CME, governs the 
    evolution of the system, we would still have the scaling laws Eqs.~\eqref{ACCscaling1} and \eqref{ACCscaling2}, 
    but the evolution nevertheless would have 
    relatively large uncertainty, 
    because of the exponential instability from the CPI. 
    For later purposes,
    we keep the ambiguity in the scaling evolution and 
    introduce a scaling index $q_5$ such that
    \begin{align}
    \aver{\mu_5}=-\aver{\mu_{\rm M}^+}+\aver{\mu_{\rm M}^-} \propto \eta^{-q_5}, \label{ACCscaling3}
    \end{align}
    which will be used in Sec.~\ref{Application}. 
\end{enumerate}
In \Tab{TAnalyticTimeScale}, we summarize the characteristic 
time scales relevant for the evolution of the system;
see \App{AdditionalScales} for a summary of additional time scales
and wave numbers defined in this paper.

Some of the features described above will be confirmed by direct numerical
simulations in \Sec{Results}.
They can have important consequences for baryon production, as will be
discussed at the end of the paper.

\section{Results}
\label{Results}

In this section, we show the results of the direct numerical simulation.
We first study the case with $k_0 \ll |\mu_{50}|$
until Sec.~\ref{sec:IIIG}. 
In Sec.~\ref{sec:IIIH} we study the case with $k_0 \gg |\mu_{50}|$.
Some of our observations will turn out to be consistent with the
theoretical prediction discussed in Sec.~\ref{IIE}.
We will also see some other features that have not been addressed there.

\subsection{Visualization of magnetic and fermion chiralities} 

We begin by discussing the simulation of Ref.~\cite{BKS23}
with $k_0 \ll |\mu_{50}|$, which we refer to as Run~O.
In \Fig{AB}, we present visual impressions of magnetic and fermion
chiralities in Run O at different times.
We see that the turbulent structures gradually grow in size and the
extreme values away from zero decrease as time goes on.
Furthermore, $\mu_5$ and $\AAA\cdot\BB\,\lambda/2$ have predominantly
opposite signs, as expected.
Locally, however, there is no correspondence between the two fields.
This is because the vanishing total chirality is only a statistical
property.

\subsection{Evolution of characteristic scales}
\label{EvolCharScales}

As discussed in Ref.~\cite{BKS23}, it is important to allow for
sufficient scale separation between the smallest available wave number
$k_1\equiv2\pi/L$ and the initial wave number of the peak, $k_0$.
It is also important that there is enough separation between $k_0$
and the initial wave number of the CPI, $|\mu_{50}|/2$, 
to confirm distinct features of the evolution of the system.
Both, $k_0$ and $|\mu_{50}|/2$, in turn, must be much smaller than the
largest available wave number $k_{\rm Ny}=k_1N/2$.
Sufficient scale separation between $k_1$ and $k_0$ is particularly
important for obtaining the theoretically expected increase of
$\xiM\propto\eta^{4/9}$ along with the decay of
$\EEM\propto\eta^{-10/9}$, based on the conservation of the
Hosking integral adapted to the total chirality.
Indeed, in Run~O, an optimized balance between the two scale separation
requirements has been achieved.

\begin{figure}[t!]\begin{center}
\includegraphics[width=\columnwidth]{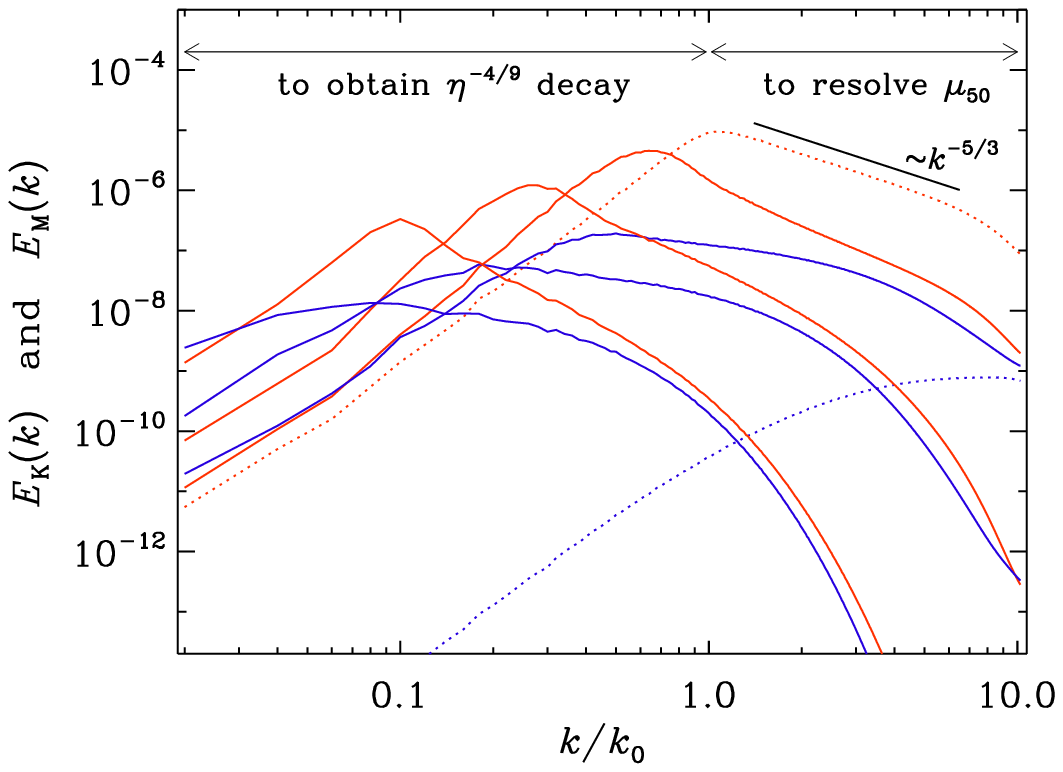}
\end{center}\caption[]{
Magnetic energy (red lines) and kinetic energy (blue lines) spectra for
Run~O at times $\eta=460$, 4600, and 46,000.
The dotted lines denote the earliest outputted time $\eta=0.3$.
The straight black line indicates the initial $k^{-5/3}$ spectrum
for the magnetic field.
In the upper part, the two-sided arrows indicate the requirements
for scale separation at small and large $k$ to obtain the
$\kp\propto\eta^{-4/9}$ decay and to resolve $|\mu_{50}|$, respectively.
}\label{rspec_select_ub_1024a_mu10_k002o}\end{figure}

With the start of the simulation, the helical random magnetic field, which
is present initially, drives turbulent motions through the Lorentz force.
This causes $\EK(k)$ to grow quickly at all wave numbers, but it is
always less than $\EM(k)$; see \Fig{rspec_select_ub_1024a_mu10_k002o},
where we compare kinetic and magnetic energy spectra at different times.
After some time, $\EK(k)$ approaches $\EM(k)$ at large values of $k$,
i.e., the motions are in approximate equipartition with the magnetic
field at high wave numbers.
This observation supports the estimate of $\urms^\mathrm{max}$
for ${\tilde v}_\lambda$; see \Eq{tildevlam}.
At the same time, the initial $k^{-5/3}$ spectrum for $\EM(k)$ develops
into a slightly steeper one due to finite magnetic diffusion.
In \Fig{rspec_select_ub_1024a_mu10_k002o}, we also mark the two scale
separation ratios.

%FIG2

\begin{figure}[t]\begin{center}
\includegraphics[width=\columnwidth]{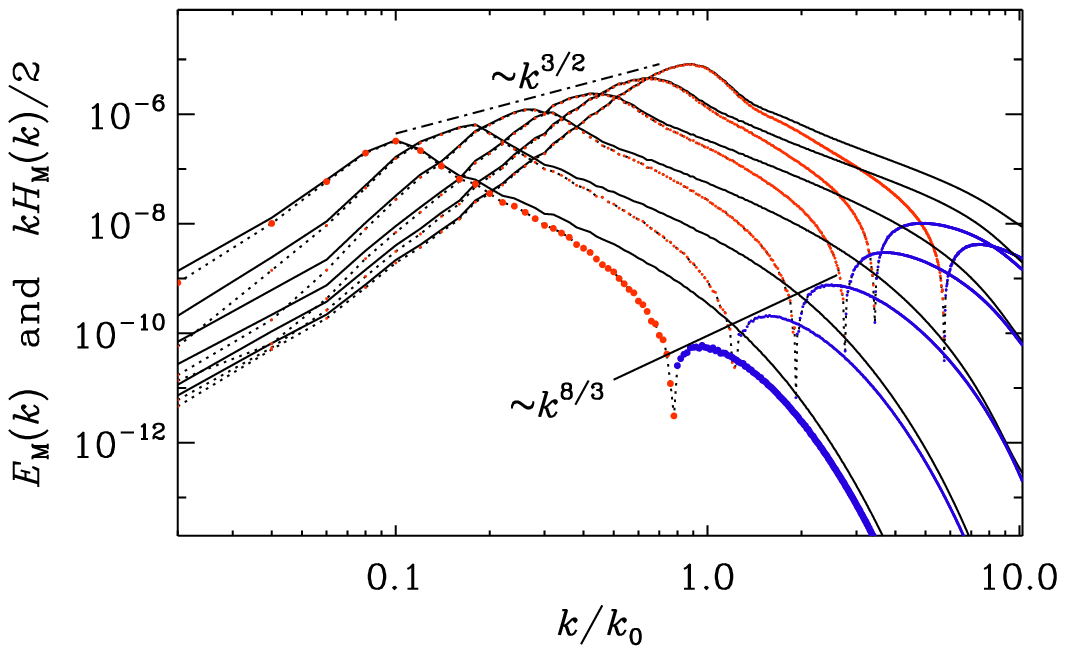}
\end{center}\caption[]{
Magnetic energy (solid lines) and normalized helicity spectra
$k\HM(k)/2$ (dotted lines with red and blue symbols for positive
and negative helicity spectra, respectively) for Run~O at
times $\eta=150$, $460$, 1500, 4600, 15,000, and 46,000.
The peaks $k_\mathrm{I}$ (peaks of the red curves) and $k_\mathrm{II}$
(peaks of the blue curves) evolve underneath the envelopes $\propto
k^{3/2}$ and $\propto k^{8/3}$, respectively.
}\label{rspec_select_1024a_mu10_k002o}\end{figure}

As already discussed in Ref.~\cite{BKS23}, even though there is vanishing
net chirality, $\aver{\mu_{\rm M}}+\aver{\mu_5}=0$, there is still some
degree of inverse cascading, just like in nonhelical magnetically dominated
turbulence \citep{BKT15,BK17}.
We see this clearly in \Fig{rspec_select_1024a_mu10_k002o}, where the position
of the magnetic peak, $k_{\rm I}(\eta)$, gradually moves to smaller values.
At the same time, the height of the peak decreases, following an
approximate power law $\propto k^\beta$, with $\beta=3/2$; see
\Fig{rspec_select_1024a_mu10_k002o}. 
This can be explained by the conservation of the Hosking integral
\cite{Hosking+Schekochihin21, Zhou+22}; see also Eq.~\eqref{hoscons}.
The exponent $\beta=3/2$ is characteristic of the fact that the net
chirality vanishes, even though near the peak itself the field is
locally fully helical, as we see from the proximity of 
$k|\HM(k)|/2$ and $\EEM(k)$; see \Eq{realizability}.

%FIG3

The newly injected magnetic helicity from the CPI leads to
a growth of the magnetic field at large wave numbers. 
It manifests itself mostly through the build-up of negative magnetic
helicity at high wave numbers.
At some point, we also see a gradual propagation of the
secondary peak $k_{\rm II}$ toward smaller $k$, 
which has not been addressed in Sec.~\ref{IIE}.
It lies underneath an envelope with an approximate $k^{8/3}$ slope; see
\Fig{rspec_select_1024a_mu10_k002o}.
At present, the exponent 8/3 is just empirical and there is no theory
for it.
It should be noted, however, that in other cases with a shorter inertial
range, we have found larger exponents.
Thus, the exponent could also be smaller when the inertial range is larger,
i.e., when there is more scale separation and $\aver{\mu_5}\xiM$ is larger.

Another characteristic wave number is $k_\pm$, where the sign of the
spectral magnetic helicity changes.
It is used in the definitions of $\aver{\mu_{\rm M}^+}$ and
$\aver{\mu_{\rm M}^-}$ in \Eqs{muMp}{muMm}.

%FIG4
\begin{figure}\begin{center}
\includegraphics[width=\columnwidth]{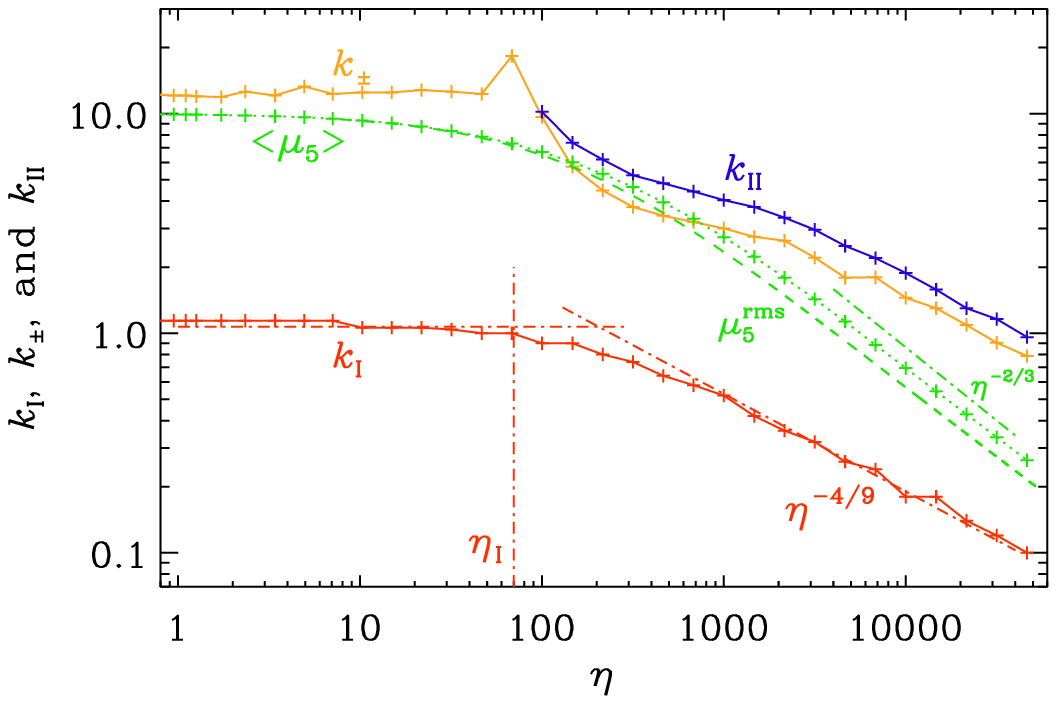}
\end{center}\caption[]{
Comparison of $k_{\rm I}$ (red), $k_\pm$ (orange),
and $k_{\rm II}$ (blue) for Run~O.
The green dashed line shows $\aver{\mu_5}$ and the
green dotted line shows the rms value $\mu_5^{\rm rms}$.
The sloping red (green) dashed-dotted line indicates
$\eta^{-4/9}$ ($\eta^{-2/3}$) scaling.
}\label{pkmax_comp_1024a_mu10_k002o}\end{figure}

%FIG5
\begin{figure}\begin{center}
\includegraphics[width=\columnwidth]{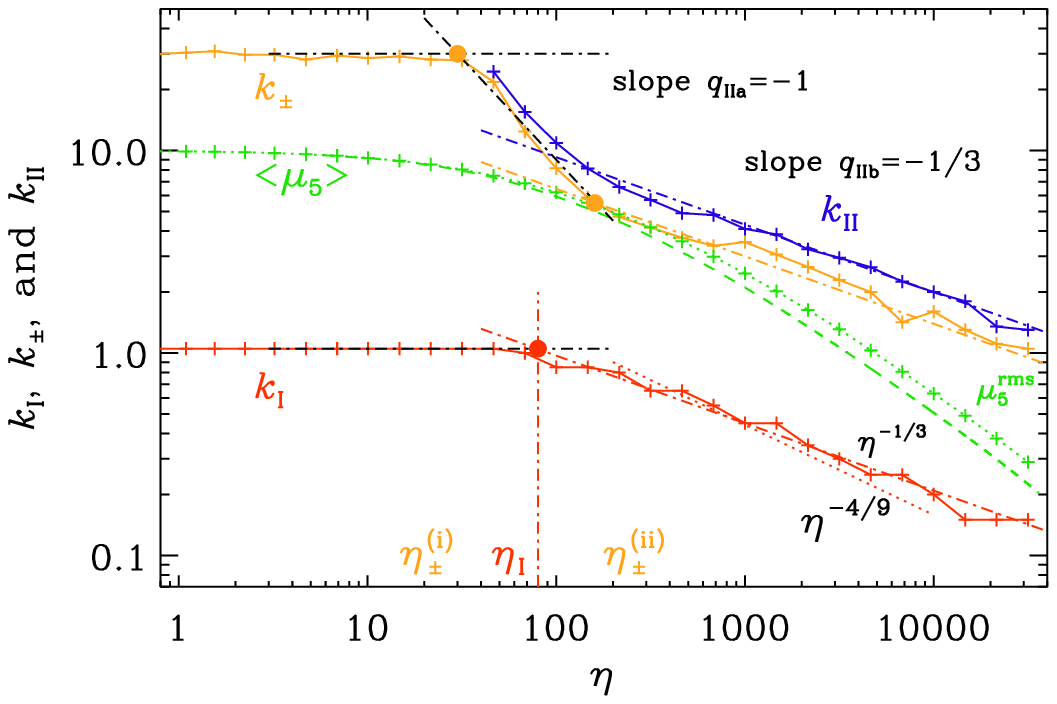}
\end{center}\caption[]{
Similarly to \Fig{pkmax_comp_1024a_mu10_k002o}, but for Run~I.
The red dashed-dotted line indicates here the $\eta^{-1/3}$ scaling,
which describes the $k_{\rm I}$ scaling better than the
$\eta^{-4/9}$ scaling indicated by the red dotted line.
The orange and red dots indicate the crossings of the extrapolated
tangents on which the times $\eta_\pm^\mathrm{(i)}$, $\eta_\mathrm{I}$,
and $\eta_\pm^\mathrm{(ii)}$ are based.
}\label{pkmax_comp_1024a_mu10_rep}\end{figure}

\begin{table*}\caption{
Summary of the runs discussed in this paper.
Except for Run~P, where $\lambda=500$, we have in all other cases $\lambda=2\times10^4$.
Runs~A and B below the last horizontal line have nonvanishing
net chirality and are discussed at the end of the paper.
The asterisk on the value of $k_1=0.01$ Run~J'' indicates that the
resolution is $N^3=2048^3$, so the Nyquist wave number is here the
same as for Run~J' with $k_1=0.02$.
}\vspace{12pt}\centerline{\begin{tabular}{lccccccccccccccccccc}
Run & $k_0$ & $-\mu_{50}$ & $\sigma^{-1}$ & SLD & $\eta_{\rm CPI}$ & $\eta_{\rm diff}$ & $\eta_{\mu_\mathrm{M}^+}$ & $q_{\rm I}$ & $\eta_{\rm flip}$ &
$\eta_\pm^\mathrm{(i)}$ & $\eta_\mathrm{I}$ & $\eta_\pm^\mathrm{(ii)}$ & $\eta_{\mu_\mathrm{M}^-}$ & $|\aver{\muM^-}|^{\max}$ &
$k_\pm/k_{\rm I}$ & $k_{\rm II}/k_{\rm I}$ & $v_\lambda$ & $\urms^{\max}$ & $k_1$ \\ % $\eta_{\max}$
\hline
VI   &  1 &160 & $5\times10^{-4}$ &  no &0.08& $2\times10^3$ &   3 & 1/3 & --- & 0.3&   45 &   14 &   7 & 43     & 1.4 & 2.0 & 1.13  & 0.095 & 0.2  \\%  5,300 \\ % 1024c_mu160_k02
V    &  1 & 80 & $5\times10^{-4}$ &  no &0.3 & $2\times10^3$ &   4 & 1/3 & --- & 0.5&   55 &   40 &   4 & 13     & 1.5 & 1.8 & 0.57  & 0.050 & 0.2  \\%  5,300 \\ % 1024c_mu80_k02
IV   &  1 & 50 & $5\times10^{-4}$ &  no &0.8 & $2\times10^3$ &   6 & 1/3 & --- & 1.6&   50 &   55 &  50 & 5.1    & 1.5 & 2.0 & 0.35  & 0.050 & 0.1  \\%  5,300 \\ % 1024c_mu50
III  &  1 & 30 & $5\times10^{-4}$ &  no &2.2 & $2\times10^3$ &  14 & 1/3 & --- & 1.7&   80 &  108 & 150 & 1.5    & 1.8 & 2.3 & 0.21  & 0.018 & 0.2  \\%  5,300 \\ % 1024c_mu30_k02b
II+  &  1 & 20 & $5\times10^{-4}$ &  no & 5  & $2\times10^3$ &  18 & 1/3 & --- &  6 &   75 &  160 & 200 & 0.46   & 1.8 & 3.1 & 0.141 & 0.031 & 0.1  \\%    550 \\ % 1024a_mu20_5em4
II   &  1 & 20 & $2\times10^{-4}$ &  no &12.5& $5\times10^3$ &  60 & 1/3 & --- &  9 &   75 &  120 & 200 & 0.009  & 4.4 & 6.7 & 0.141 & 0.032 & 0.1  \\%  5,300 \\ % 1024a_mu20
II$-$&  1 & 20 & $1\times10^{-4}$ &  no & 25 & $1\times10^4$ & 160 & 1/3 & --- & 14 &   75 &  140 & 200 & 0.003  & 6.7 & 10  & 0.141 & 0.031 & 0.1  \\%  1,100 \\ % 1024b_mu20_1em4
I    &  1 & 10 & $2\times10^{-4}$ &  no & 50 & $5\times10^3$ & 125 & 1/3 & --- & 30 &   80 &  160 & 250 & 0.009  & 6.7 & 9.6 & 0.071 & 0.030 & 0.05 \\% 36,000 \\ % 1024a_mu10_rep
O    &  1 & 10 & $2\times10^{-4}$ &  no & 50 & $5\times10^3$ & 125 & 4/9 & --- & 20 &   70 &  120 & 300 & 0.008  & 8.1 & 9.6 & 0.071 & 0.0123& 0.02 \\% 52,000 \\ % 1024a_mu10_k002o
O'   &  1 & 10 & $2\times10^{-4}$ & yes & 50 & $5\times10^3$ & 125 & 4/9 & --- & 20 &  110 &  120 & 400 & 0.015  & 6.4 & 8.4 & 0.071 & 0.0103& 0.02 \\% 34,000 \\ % 1024a_mu10_k002b
L    &  1 & 10 & $2\times10^{-4}$ & yes & 50 & $5\times10^3$ & 125 & 4/9 & --- &180 &  400 &  500 &1000 & 0.027  & 6.7 & 8.7 & 0.071 & 0.0079& 0.01 \\%147,000 \\ % 1024a_mu10_k001c
M    &  1 &  7 & $2\times10^{-4}$ & yes &102 & $5\times10^3$ & 165 & 1/3 & --- &260 &  220 &  800 & 800 & 0.006  & 5.8 & 7.2 & 0.049 & 0.0065 & 0.01 \\%  6,000 \\ % 1024a_mu7_k001a
N    &  1 &  5 & $2\times10^{-4}$ & yes &200 & $5\times10^3$ & 235 & 1/3 & --- &350 &  200 &  800 &1000 & 0.0015 & 6.3 & 7.8 & 0.035 & 0.0055 & 0.01 \\%  6,000 \\ % 1024a_mu5_k001a
N'   &  1 &  5 & $2\times10^{-4}$ & yes &200 & $5\times10^3$ & 200 & 4/9 & --- &--- &  800 & 3000 &15000& 0.00004& 5.5 & 2.4 & 0.035 & 0.0035& 0.005\\% 49,000 \\ % 1024a_mu5_k0005b
F    &  1 &  5 & $2\times10^{-4}$ & yes &200 & $5\times10^3$ & --- & 4/9 & 100 &--- &  250 & 9000 & --- & ---    & 30  & 30  & 0.035 & 0.0055& 0.01 \\% 23,000 \\ % 1024d_mu5_Gam1em2_lam2e4
J    &  1 &  5 & $5\times10^{-4}$ &  no & 80 & $2\times10^3$ &  71 & 4/9 & --- &230 &  300 &  500 & 700 & 0.0003 & 6.1 & 7.6 & 0.035 & 0.0068 & 0.01 \\%  6,800 \\ % 1024b_mu5_k001a
J''  &  1 &  5 & $5\times10^{-4}$ &  no & 80 & $2\times10^3$ &  71 & 4/9 & --- & 90 &  300 &  500 & 460 & 0.0006 & 6.1 & 7.6 & 0.035 & 0.0071 & 0.01*\\%  3,000 \\ % 2048a_mu5_k001b
J'   &  1 &  5 & $5\times10^{-4}$ &  no & 80 & $2\times10^3$ &  76 & 1/3 & --- & 95 &  120 &  500 & 460 & 0.0005 & 5.8 & 7.7 & 0.035 & 0.0070 & 0.02 \\%  2,000 \\ % 1024a_mu5_k002b
P    &  1 & 0.1& $2\times10^{-4}$ &  no &$5\times10^5$& $5\times10^3$ &$10^4$&3/5 & --- & -- &  160 &  --- & --- &$3\times10^{-9}$ & --- & --- & 0.001 &0.0070 & 0.02 \\% 1024a_mu01_k002o <<==
G    & 0.5& 10 & $2\times10^{-4}$ &  no & 50 & $2\times10^4$ & 200 & 1/3 & --- & 30 &  360 &  360 & 300 & 0.044  & 4.6 & 6.6 & 0.071 & 0.0188 & 0.05 \\%  7,600 \\ % 1024a_mu10_k05_1024proc2
H    & 0.2& 10 & $2\times10^{-4}$ &  no & 50 &$1.2\times10^5$& 375 & 1/3 & --- & 75 & 2000 & 2000 &3000 & 0.42   & 2.0 & 2.7 & 0.071 & 0.0074 & 0.02 \\% 20,000 \\ % 1024b_mu10_k02_1024proc
\hline
A    &  1 & 10 & $2\times10^{-4}$ &  no & 50 & $5\times10^3$ & --- & 4/9 & --- &  8 &  110 &  210 & --- & ---    & 7.3 & 9.5 & 0.071 & 0.0109 & 0.02 \\% 52,000 \\ % 1024a_mu10_k002o_less
B    &  1 & 10 & $2\times10^{-4}$ &  no & 50 & $5\times10^3$ & --- & 4/9 & --- & 20 &   90 &  120 & 250 & ---    & 7.8 & 8.9 & 0.071 & 0.0137 & 0.02 \\% 52,000 \\ % 1024a_mu10_k002o_more
\label{Ttimescales}\end{tabular}}\end{table*}
% data/axel/chiral_fluids/turbulent_decay/idl
% scr/axel/chiral_fluids/turbulent_decay/idl
% cvs co axel/chiral_fluids/turbulent_decay/1024a_mu10_k002o
% mv !$ .

In \Fig{pkmax_comp_1024a_mu10_k002o}, we plot the evolution of the
characteristic wave numbers $k_{\rm I}$, $k_\pm$, and $k_{\rm II}$.
We clearly see the $k_{\rm I}\propto\eta^{-4/9}$ decay predicted by
the conservation of the Hosking integral adapted to the total chirality
\cite{BKS23}.
It emerges after a time $\eta_{\rm I}$, which is expected to be close to
$\eta_\lambda$ (and also $\eta_\mathrm{turb}$); see Eq.~\eqref{etalam}.
In Run~O we find $\eta_{\rm I}\approx100$.

The evolution of $k_\pm$ and $k_{\rm II}$ can be seen more clearly
when the Nyquist wave number is larger.
We therefore discuss in \Fig{pkmax_comp_1024a_mu10_rep} another run, also
with $N=1024^3$ mesh points, but now with $k_1=0.05$ (instead of 0.02),
so $k_{\rm Ny}=25.6$, which is a little over five times larger than
$|\mu_{\rm 50}|/2=5$.
In \Tab{Ttimescales}, this run is referred to as Run~I, which differs
from the previously discussed Run~O mainly in the value of $k_1$.
It also has a shallower scaling of the correlation length,
$\xiM\propto k_\mathrm{I}^{-1} \propto \eta^{1/3}$, which seems to be an artifact caused by
insufficient scale separation, i.e., the value of $k_1$ is not
sufficiently small.
Empirically, we find that, if $k_0/k_1 \gg 20$, there is an
inverse cascade with $\xiM \propto k_\mathrm{I}^{-1} \propto \eta^{4/9}$.
The parameters $\eta_\mathrm{I}$, $\eta_\pm^\mathrm{(i)}$, and
$\eta_\pm^\mathrm{(ii)}$, listed in \Tab{Ttimescales}, are discussed
below.
We also give here the values of $v_\mu$ and $v_\lambda/v_\mu$.
Run~O' is similar to Run~O, except that here, SLD has been added.
The two runs are virtually indistinguishable.

The evolution of the peaks of the spectrum can be summarized as follows. 
(i) After the start of the run, the CPI induces a growth of the negative
helicity modes at the secondary peak $k_\mathrm{II}$, which stays constant
until $\eta=\eta_\pm^\mathrm{(i)}$, and then starts to decrease with time in a
power law fashion, $k_\mathrm{II} \propto\eta^{-q_{\rm IIa}}$, with
$q_{\rm IIa}\approx1$ in all cases.
(ii) The original large-scale spectrum is unchanged until some time
$\eta=\eta_{\rm I}$ and then starts to decrease via an inverse cascade with
$k_\mathrm{I}(\eta) \propto\eta^{-q_{\rm I}}$, where $q_{\rm I}$ is
expected to be equal to the exponent $q=4/9$ found in Ref.~\cite{BKS23}.
(iii) At time $\eta=\eta_\pm^\mathrm{(ii)}$, the decay of
the secondary peak becomes slower with a smaller index,
$k_\mathrm{II} \propto\eta^{-q_{\rm IIb}}$, with
$q_{\rm IIb}<q_{\rm IIa}\approx1$.
Those parameters are summarized in \Tab{Ttimescales}.

The plot of characteristic wave numbers $k_{\rm I}$, $k_\pm$, and
$k_{\rm II}$ in \Fig{pkmax_comp_1024a_mu10_rep} shows three distinct
times, $\eta_\pm^{\rm(i)}\la\eta_{\rm I}\la\eta_\pm^{\rm(ii)}$, where
$k_\pm$ begins to decrease first rapidly, at $\eta=\eta_\pm^{\rm(i)}$,
and later, at $\eta=\eta_\pm^{\rm(ii)}$, more slowly, approximately like
$\eta^{-4/9}$, i.e., $q_{\rm IIb}\approx q=4/9$.
The decay of $k_{\rm II}$ closely follows that of $k_\pm$.
The decay of $k_{\rm I}$, on the other hand, does not show the rapid
decay phase that we see in $k_\pm$ and $k_{\rm II}$, but turns directly
into the approximate $\eta^{-4/9}$ decay at $\eta=\eta_\pm^{\rm(i)}$.

\subsection{Onset of inverse cascading}

\begin{figure}\begin{center}
\includegraphics[width=\columnwidth]{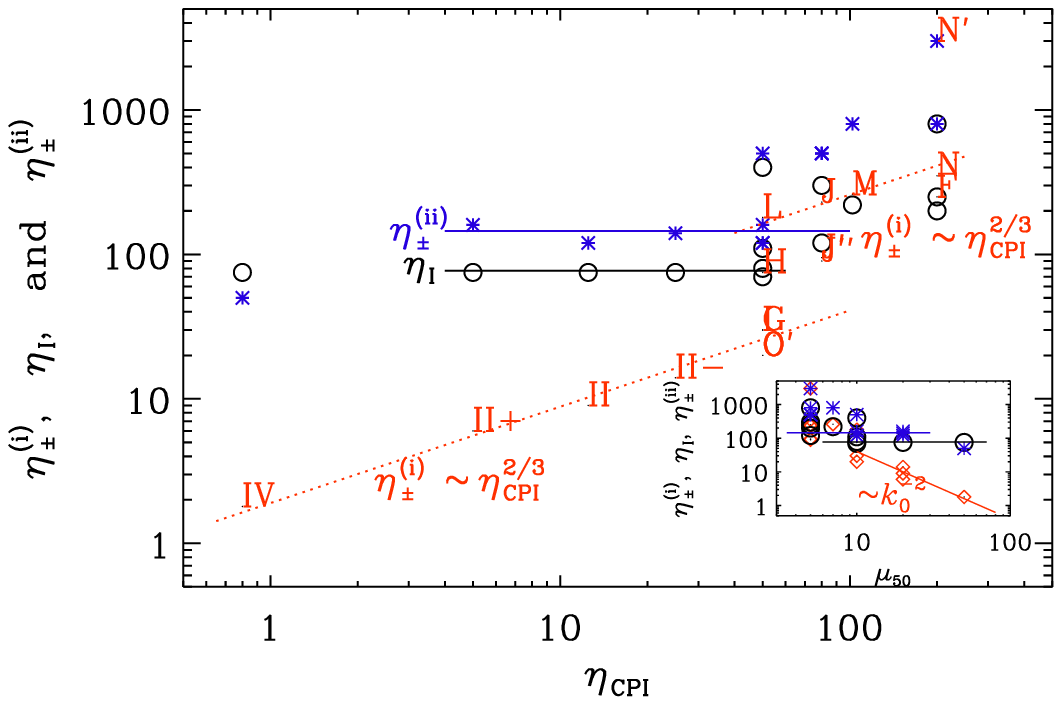}
\end{center}\caption[]{
Dependence of $\eta_\pm^{\rm(i)}$, $\eta_{\rm I}$, and
$\eta_\pm^{\rm(ii)}$ on $\eta_{\rm CPI}$.
$\eta_\pm^{\rm(i)}$ shows an approximate $\eta_{\rm CPI}^{2/3}$
dependence along two branches that are separated by a factor of about 6.
$\eta_{\rm I}$ and $\eta_\pm^{\rm(ii)}$ are essentially
independent of $\eta_{\rm CPI}$.
The inset shows that $\eta_\pm^{\rm(i)}$ scales inverse
quadratically with $|\mu_{50}|$.
}\label{peta1_vs_mu50}\end{figure}

It is of interest to vary the separation between $|\mu_{50}|/2$ and $k_0$
to see the dependence of the relevant characteristic times on these
wave numbers.
We have performed simulations for different values and consider runs
%where we change $k_{\rm I}$ and keep $\mu_{50}$ fixed, and others where
%AB: k_I -> k_0
where we change $k_0$ and keep $\mu_{50}$ fixed, and others where
we change $\mu_{50}$ and keep $k_0$ fixed.
It both cases, of course, since we want to satisfy
$\aver{\mu_5}+\aver{\mu_{\rm M}}=\const$, we need to adjust the
amplitude of the initial magnetic field correspondingly.
The results are summarized in \Tab{Ttimescales} and plotted
in \Figs{peta1_vs_mu50}{peta2_vs_vlam}.

%FIG6
%/home/brandenb/tex/mhd/kohei/idl

%FIG7
\begin{figure}\begin{center}
\includegraphics[width=\columnwidth]{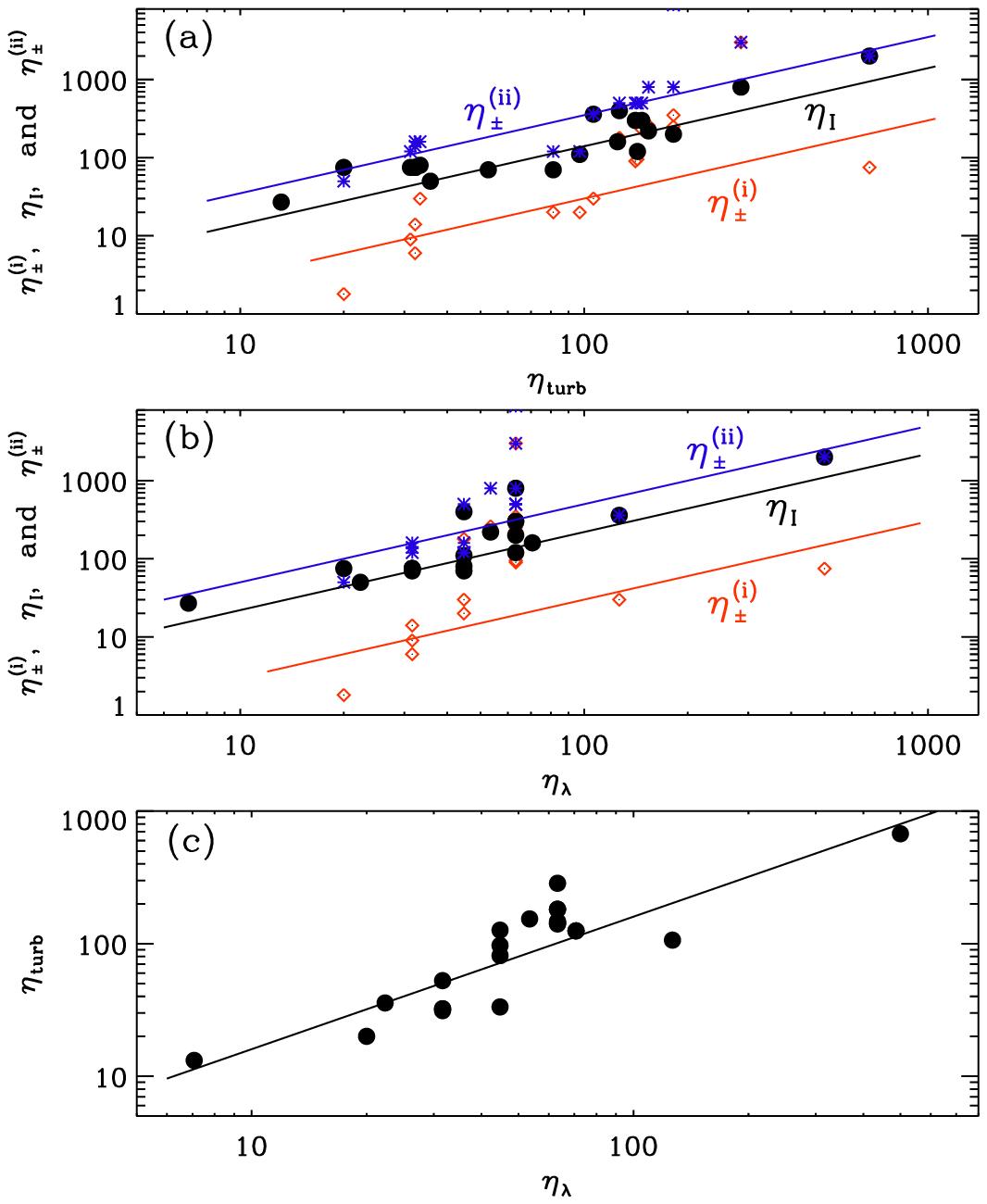}
\end{center}\caption[]{
Dependence of $\eta_{\rm I}\approx\eta_\pm^\mathrm{(ii)}$ and
$\eta_\pm^\mathrm{(i)}$ on (a) $\eta_{\rm turb}$ and
(b) $\eta_\lambda$, as well as (c) the dependence of
$\eta_{\rm turb}$ on $\eta_\lambda$.
}\label{peta2_vs_vlam}\end{figure}

One may presume that $\eta_\pm^{\mathrm{(i)}}$ is roughly estimated by
$\eta_\mathrm{CPI}$ since the grow of negative helicity modes becomes
effective at that time.
We see, however, that, while $\eta_\pm^{\rm(i)}$ decreases
quadratically with increasing $|\mu_{50}|$, the dependence on
$\eta_{\rm CPI}=\sigma\mu_{50}^{-2}$ is shallower than linear and follows
approximately an $\eta_{\rm CPI}^{2/3}$ scaling; see \Fig{peta1_vs_mu50}.
Thus, $k_{\rm II}$ starts to decline more rapidly when $|\mu_{50}|$ is
large, although it is unclear why this exponent is here $\approx2/3$.
On the other hand, we see that the five data points with $k_1=0.01$
(Runs~L, M, N, J, and J'' with smaller $|\mu_{50}|$) lie on another
$\eta_{\rm CPI}^{2/3}$ line that is shifted upward by a factor of about
6 relative to the runs with larger $k_1$.
The reason for this is that for large values of $\eta_{\rm CPI}$, it
became necessary to decrease the value of $k_1$.
This decreased the Nyquist wave number since $N$
remained unchanged, which can cause artifacts in the values of $k_\pm$.
Small values of $k_1$ also facilitates the $\eta^{4/9}$
scaling of $\xiM$ and related length scales; see the comparison between
Runs~N and N' in \Tab{Ttimescales}.
This shows that $\eta_\pm^{\rm(i)}$ is currently very sensitive to
these restrictions which will be alleviated in future with larger
computational power.
Nevertheless, there is clearly a trend for an uprise in the dependence
of $\eta_\pm^{\rm(i)}$ on $\eta_{\rm CPI}$ for large values.

Next, we examine the dependence of $\eta_\mathrm{I}$ and
$\eta_\pm^\mathrm{(ii)}$ on $k_0$ and $\mu_{50}$.
Figure~\ref{peta1_vs_mu50} shows that
the time $\eta_{\rm I}$ of the onset of the decline of $k_{\rm I}$
does not strongly depend on the value of $\mu_{50}$.
Likewise, the time $\eta_\pm^\mathrm{(ii)}$ when the decay of $k_{\rm II}$ slows
down, does not strongly depend on $\mu_{50}$.
Again, however, there is an upward shift of data points for the four runs,
for which $k_1=0.01$.
As discussed in Sec.~\ref{IIE}, we expect that $\eta_\mathrm{I}$
is close to $\eta_\mathrm{turb}$ and $\eta_\lambda$. 
The upper two panels of Fig.~\ref{peta2_vs_vlam} show the dependence 
of $\eta_\pm^{\rm(i)}$, $\eta_{\rm I}$, and $\eta_\pm^{\rm(ii)}$ on $\eta_\mathrm{turb}$ and $\eta_\lambda$,
respectively.
From these plots, we estimate that
\begin{equation}
\eta_{\rm I}\approx1.4\, \eta_{\rm turb} \approx 2.2\, \eta_\lambda.
\label{etaIscaling}
\end{equation}
In the lowest panel of Fig.~\ref{peta2_vs_vlam}, we also show the relation
between $\eta_{\rm turb}$ and $\eta_\lambda$, i.e.,
\begin{equation}
    \eta_\mathrm{turb} \approx  1.6\, \eta_\lambda, 
\end{equation}
which shows the validity of the estimate of $\urms^\mathrm{max}$ 
in terms of ${\tilde v}_\lambda$. 
Equation~\eqref{etaIscaling} is useful for estimating the properties of
magnetic field strength and coherence length at later times.
Therefore, we conclude that the numerical results support,
at least for a moderate scale separation, $1 < |\mu_{50}|/k_0 \lesssim {\cal O}(10)$,
the theoretical estimate for the evolution of the characteristic scales
given in Sec.~\ref{IIE} with a more accurate determination 
of the time of the onset of the scaling evolution, \Eq{etaIscaling}.

%FIG8
\begin{figure}\begin{center}
\includegraphics[width=\columnwidth]{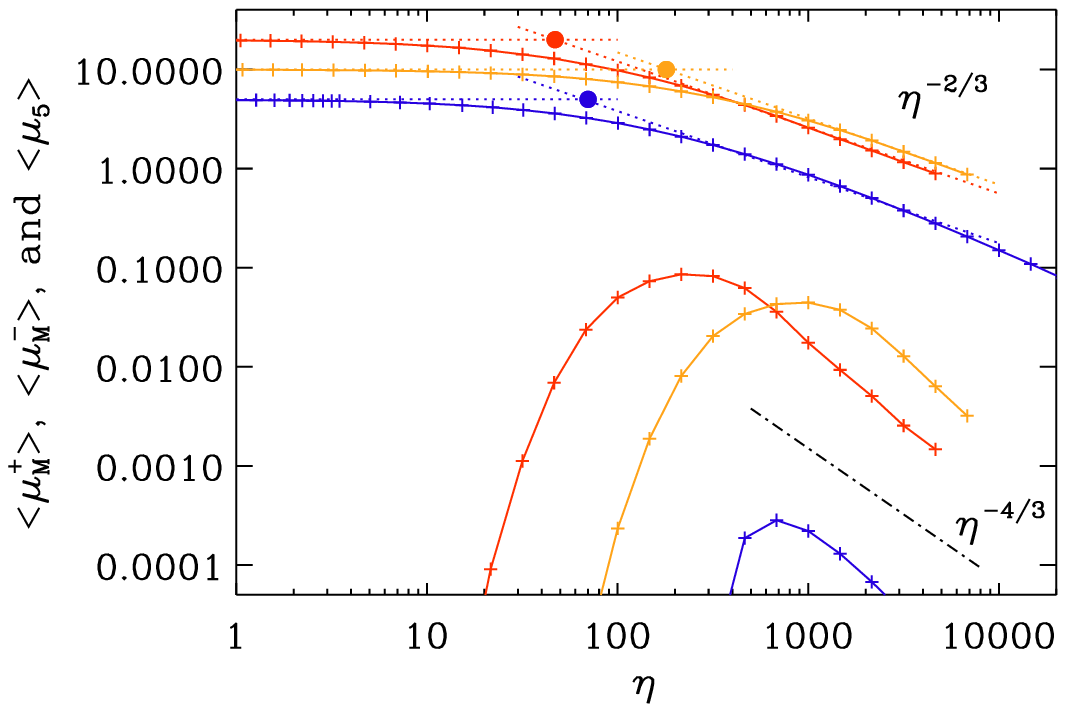}
\end{center}\caption[]{
Comparison of $\aver{\mu_{\rm M}^\pm}$ for Run~II (red lines),
Run~J (blue lines), and Run~G (orange lines).
The times $\eta_{\rm ACC}$ are marked by the correspondingly
colored filled symbol at the crossing points of the extrapolated
$\eta^{-2/3}$ decay law with the initially constant values,
indicated by dotted line.
The $\eta^{-4/3}$ decay law $\aver{\muM^-}$ is shown as the
dashed-dotted line.
}\label{phmax_comp}\end{figure}

\subsection{Evolution of $\aver{\mu_5}$ and $\aver{\mu_{\rm M}^\pm}$}
\label{subsec:evolution}

We now discuss how the chirality of the system evolves.
Using \Eqs{muMp}{muMm}, we divide the magnetic helicity into
$\aver{\mu_{\rm M}^+}$ and $\aver{\mu_{\rm M}^-}$.
The typical evolution of $\langle \mu_5 \rangle$ and
$\langle \muM^{\pm} \rangle$ is as follows.
(i) $\aver{\mu_5}$ and $\aver{\muM^{+}}$ stay constant until the time
$\eta=\eta_{\mu_\mathrm{M}^+}$, when the ACC commences
exhibiting a power law decay.
(ii) $\aver{\muM^{-}}$ grows until the time
$\eta = \eta_{\mu_\mathrm{M}^-}$ and then decays.
Thus, $\eta_{\mu_\mathrm{M}^-}$ is determined as the time
when $\aver{\muM^{-}}$ is maximum.

As discussed in Sec.~\ref{IIE}, the decay of $\aver{\mu_5}$ and
$\aver{\muM^{+}}$ due to the ACC is expected to be like $\eta^{-2/3}$.
In \Fig{phmax_comp}, we have overplotted the asymptotic $\eta^{-2/3}$
decay laws of magnetic helicity with results of some of the representative
numerical runs (Runs~II, J, and G), which clearly shows that the numerical
results support the theoretical prediction.

The decay of $\langle \muM^{-} \rangle$ is faster than that of
$\aver{\mu_5}$ and $\langle \muM^{+} \rangle$ and follows an approximate
$\eta^{-4/3}$ law, resulting in a decay of the ratio $\langle \muM^{-}
\rangle/\langle \muM^{+} \rangle\propto\eta^{-2/3}$.
Therefore, unless $\langle \muM^{-} \rangle$ becomes comparable to
$\langle \muM^{+} \rangle$ when the grow stops, a complete cancellation
between $\langle \muM^{-} \rangle$ and $\langle \muM^{+} \rangle$
never occurs.

The production of $\aver{\muM^-}$ is expected to be a result of the CPI.
We now address the question of how much $\aver{\muM^-}$ is being
produced and what its maximum value depends on.
\FFig{phmax_comp} shows that $\aver{\muM^-}$ is generally rather small,
and at least for $\mu_{50}/k_0 \lesssim 20$ there is always a strong
imbalance between $|\aver{\muM^+}|$ and $|\aver{\muM^-}|$, which never
enters a phase with a near-complete cancellation.

To see whether this is related to the value of the conductivity, we
compare simulations with different values of $\sigma$.
It turns out that runs with smaller magnetic diffusivity
($\sigma^{-1}=10^{-4}$) result in an even larger imbalance, while those
with a larger diffusivity ($\sigma^{-1}=5\times10^{-4}$) have a smaller
imbalance; compare Runs~II$+$, II, and II$-$ in \Tab{Ttimescales}.

Before closing this section, let us comment on another trend
in the numerical runs we conducted regarding the absence
of a near-complete cancellation between
$\langle \muM^{-} \rangle$ and $\langle \muM^{+} \rangle$.
For Runs~III--VI, the ratio $|\aver{\muM^-}|^{\max}/|\mu_{50}|$ becomes rather large.
This could be due to the very large scale separation of $k_0$ and $|\mu_{50}|$. 
This suggests a possibility that the CPI completes the cancellation 
between the magnetic helicity and chirality immediately. 
However, the positive and negative helicity modes are 
distributed at separate length scales with the negative ones sitting 
at higher length scales and the latter receives a stronger magnetic diffusion. 
Therefore we expect the cancellation not to be complete and that the two helicity modes decay 
with a power-law decay, not an exponential one, 
though the scaling index can be different from $-2/3$. 
In order to investigate the evolution of the system in such extreme cases, 
$|\mu_{50}|/k_0 \gg {\cal O}(10)$, 
we need to have a sufficiently large box size to realize the corresponding 
scale separation. 
The detailed study is left for future study.

\subsection{Onset of ACC}

\begin{figure}\begin{center}
\includegraphics[width=\columnwidth]{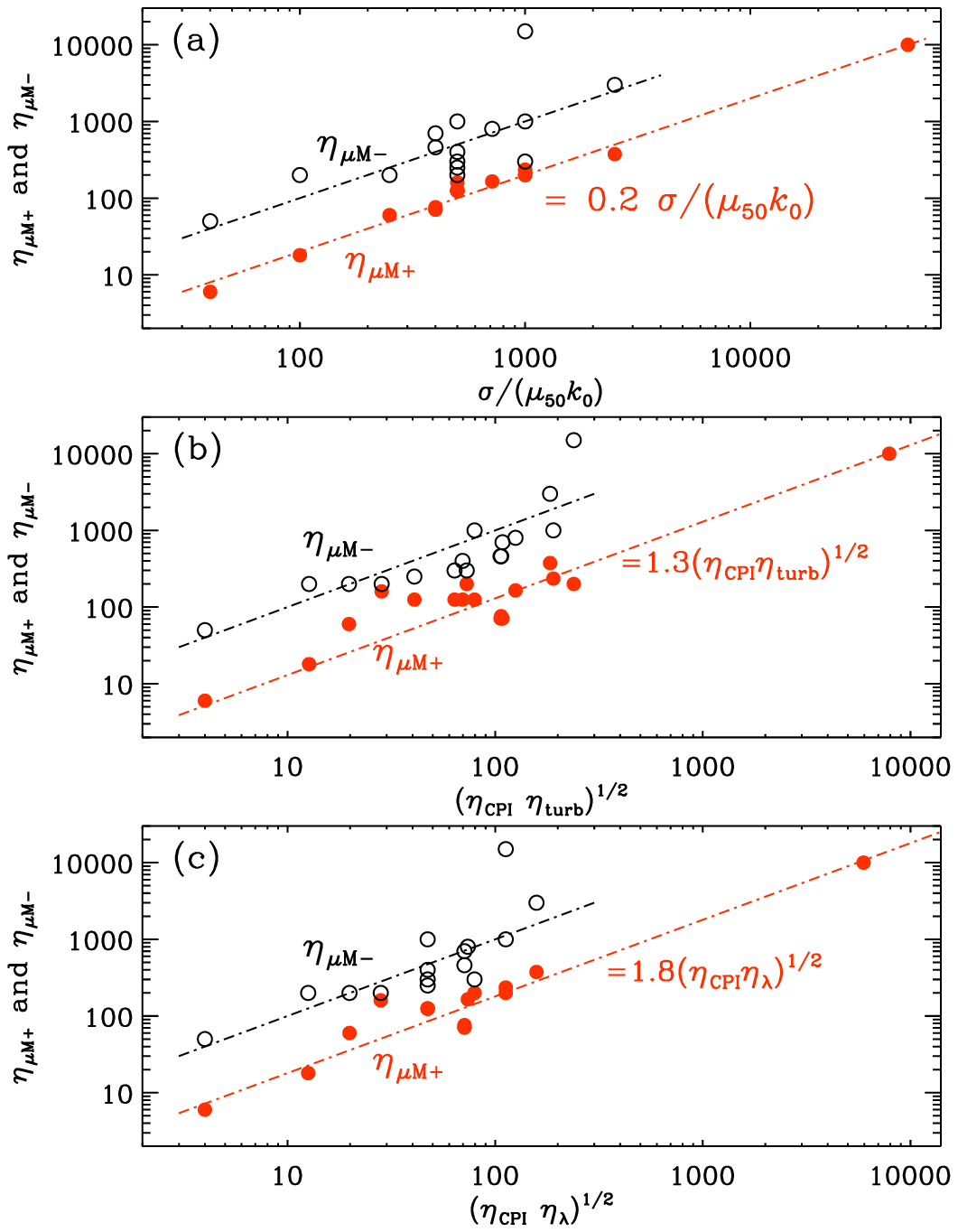}
\end{center}\caption[]{
Dependence of $\eta_{\mu_\mathrm{M}^+}$ and $\eta_{\mu_\mathrm{M}^-}$
on (a) $\sigma/|\mu_{50}k_0|$ as well as the geometric means of
$\eta_{\rm CPI}$ and (b) either $\eta_{\rm turb}$ or (c) $\eta_\lambda$.
}\label{pkpm_vs_eta45b}\end{figure}

In \Fig{pkpm_vs_eta45b}, we show the dependence of
$\eta_{\mu_\mathrm{M}^+}$ and $\eta_{\mu_\mathrm{M}^-}$ on
$\eta_\mathrm{ACC} = \sigma/|\mu_{50}k_0|$
(for the case $k_0 \ll |\mu_{50}|$; see \Eq{etaACCcase1}).
It turns out that $\eta_{\mu_\mathrm{M}^+}$ increases with
$\eta_\mathrm{ACC} = \sigma/|\mu_{50}k_0|$ such that
\begin{equation}
\eta_{\mu_\mathrm{M}^+}\approx0.2\,\eta_{\rm ACC} = 0.2\, \sigma/|\mu_{50}k_0|
\label{eta_ACC}
\end{equation}
provides a good description to the data, which supports the discussion
in Sec.~\ref{IIE}, at least for a mild hierarchy $|\mu_{50}| \gtrsim k_0$.
Furthermore, $\eta_{\mu_\mathrm{M}^-}$ shows an approximately
linear dependence on $\eta_{\rm ACC}$.
This is reasonable because the CPI becomes ineffective when the ACC
onsets such that $\aver{\mu_\mathrm{M}^-}$ is no longer amplified by
the CPI after that.

%FIG9

\subsection{The scale ratios $k_\pm/k_{\rm I}$ and $k_{\rm II}/k_{\rm I}$}

We also mention another observation in the case with $k_0 \ll |\mu_{50}|$.
At late times, the scale ratios $k_\pm/k_{\rm I}$ and $k_{\rm II}/k_{\rm I}$
reach values that are approximately constant in time.
It is about $10$ in the case of Run~O, i.e., equal to the initial scale
separation, $|\mu_{50}|/k_0=10$.
One might have expected the scale ratios to increase with $|\mu_{50}|/k_0$.
However, in all other cases, this ratio is smaller.
Some of this might also be caused by one of the two scale separation
constraints not being well enough obeyed, although the counter-intuitive
trend remains surprising.

%FIG10
\begin{figure}\begin{center}
\includegraphics[width=\columnwidth]{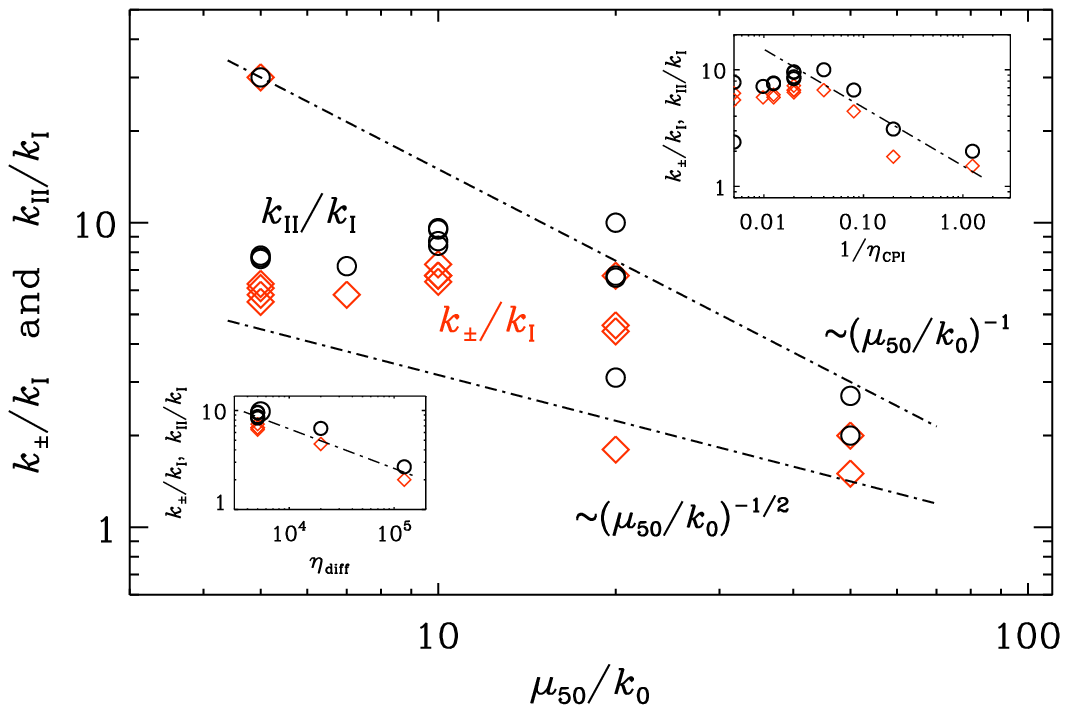}
\end{center}\caption[]{
Dependence of $k_\pm/k_{\rm I}$ and $k_{\rm II}/k_{\rm I}$
on $\eta_{\rm CPI}$, showing an $\eta_{\rm CPI}^{-0.4}$ behavior.
on $\eta_{\rm diff}$, showing a $\propto\eta_{\rm diff}^{1/2}$ behavior.
}\label{pkpm_vs_tratio}\end{figure}

In \Fig{pkpm_vs_tratio}, we show the ratios
$k_\pm/k_{\rm I}$ and $k_{\rm II}/k_{\rm I}$ versus
$|\mu_{50}|/k_0=(\eta_{\rm diff}/\eta_{\rm CPI})^{1/2}$.
The two insets give separately the dependencies on
$1/\eta_{\rm CPI}$, showing an $\eta_{\rm CPI}^{-0.4}$ behavior, and
on $\eta_{\rm diff}$, with a $\propto\eta_{\rm diff}^{1/2}$ behavior.
We see that $k_\pm/k_{\rm I}$ and $k_{\rm II}/k_{\rm I}$ decrease both
with $1/\eta_{\rm CPI}$ and with $\eta_{\rm diff}$, giving a combined
dependence on just the ratio $|\mu_{50}|/k_0$.
Thus, we see that, somewhat unexpectedly, large $|\mu_{50}|$ and small
$k_0$ tend to be detrimental to producing large scale ratios.

%FIG11
\begin{figure}\begin{center}
\includegraphics[width=\columnwidth]{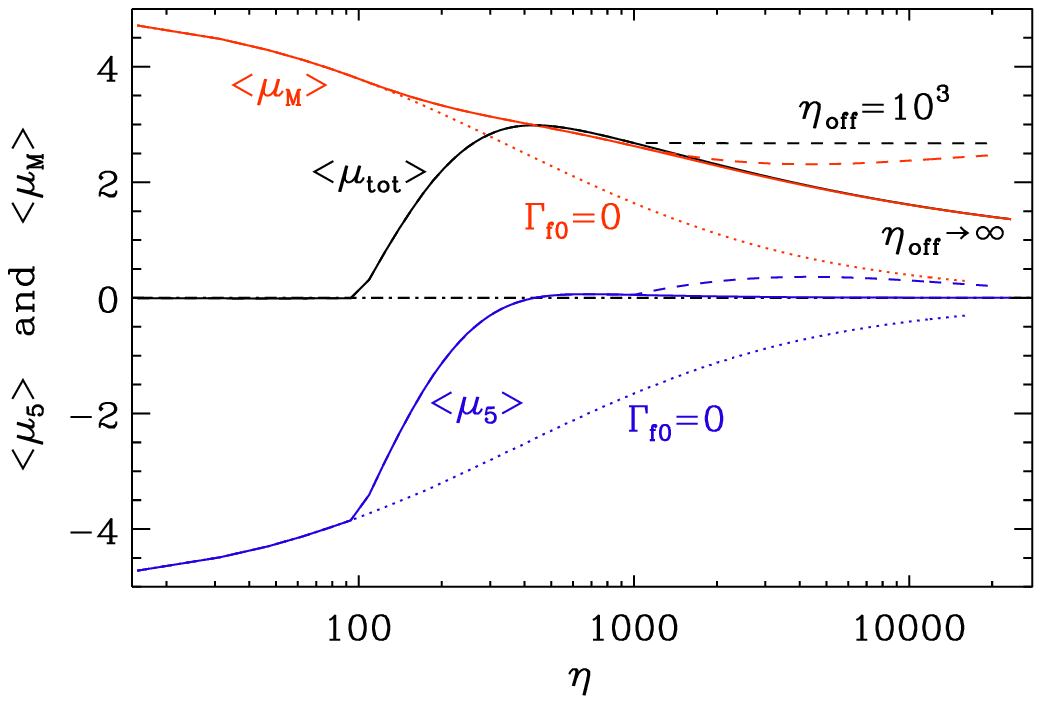}
\end{center}\caption[]{
Evolution of $\aver{\mu_{\rm M}}$ (red), $\aver{\mu_5}$ (blue), and their
sum (black) for Run~F with $\eta_{\rm flip}=100$ and $\Gamma_{\rm f0}=10^{-2}$
either for the rest of the run or only until $\eta_{\rm off}=10^3$.
The dotted lines correspond to Run~N without spin flipping.
}\label{p_comp2}\end{figure}

\subsection{Effect of chirality-flipping} \label{sec:IIIG}

The simulations discussed so far had $\Gamma=0$ and they resulted in
a final state where $\aver{\mu_5}$ and $\aver{\mu_{\rm M}}$ vanish at
late times.
As discussed in the introduction, spin flipping could prematurely
lead to a vanishing $\aver{\mu_5}$, which would imply that the decay
of $\aver{\mu_{\rm M}}$ would slow down and level off at a value away
from zero.
To study this quantitatively, we show in \Fig{p_comp2} the evolution of
$\aver{\mu_5}$, $\aver{\mu_{\rm M}}$, and $\aver{\mu_5}+\aver{\mu_{\rm M}}$
for Run~F with $\eta_{\rm flip}=100$ and $\Gamma_{\rm f0}=10^{-2}$ either for
the rest of the run or only until $\eta_{\rm off}=10^3$ (Run~F in
\Tab{Ttimescales}).
We expect that if chirality flipping becomes effective 
much later than the ACC, we do not see the effect of chirality flipping, 
while if it becomes effective much earlier, we do not see the ACC. 
This is demonstrated in \App{OthersSpinFlip}, where we show spin-flipping
versions of Runs~VI and P, where $\eta_{\rm ACC}\ll\eta_{\rm flip}$
and $\eta_{\rm ACC}\gg\eta_{\rm flip}$, respectively.
Thus we choose the parameter such that the chirality flipping 
becomes effective around (or somewhat before) the time of onset of the
ACC to capture the behavior of the system at an intermediate stage.

First, we study a case where spin flipping acts permanently (after $\eta=\eta_{\rm flip}$),
which is shown in \Fig{p_comp2} as solid lines.
We see that $|\aver{\mu_5}|$ begins to decrease rapidly to zero
after $\eta_{\rm flip}=100$.
This slows down the decay of $\aver{\mu_{\rm M}}$, which then declines
at a much smaller rate; compare with the evolution for Run~N,
which is similar to Run~F, but without spin flipping.
Qualitatively similar behaviors are also seen for smaller values of
$\Gamma_{\rm f0}$.
In all cases, we see that $\aver{\mu_5}+\aver{\mu_{\rm M}}$ evolves away
from zero.
This is because the total chirality is then no longer conserved.
The decay of $\aver{\mu_{\rm M}}$ is understood by magnetic diffusion. 
Thus we expect the decrease to slow down for a larger scale separation
between the magnetic diffusion scale and $k_{\rm I}$.

Next, it is also of interest to study a case 
where spin flipping acts 
only for a certain time interval and is then turned off again
at $\eta=\eta_{\rm off}$.
This case is shown in \Fig{p_comp2} as dashed lines.
We see that, when $\Gamma=0$ after $\eta_{\rm off}=10^3$, the sum
$\aver{\mu_5}+\aver{\mu_{\rm M}}$ is strictly constant and away from zero. 
This is in contrast to the case with permanently nonvanishing $\Gamma_{\rm f0}$,
where the sum continues to decrease slowly.
The constancy of the total chirality leads to the behavior
that $\aver{\mu_{\rm M}}$ stops to decline rapidly at a larger value.
Furthermore, during that time, some of the magnetic helicity 
decays due to the magnetic diffusion and is temporarily
converted back into fermion chirality through the total chirality conservation;
see the small increase of $\aver{\mu_5}$
with a positive maximum at $\eta\approx4000$ in \Fig{p_comp2}.
Later, however, this excess fermion chirality gets converted back into
magnetic fields, which explains the slight uprise of $\aver{\mu_{\rm M}}$
near the end of the simulation.
Indeed, this process is similar to the one seen in
Refs.~\cite{Hirono+15,Schober+2020}.
This is natural because after the decay of $\aver{\mu_5}$ the setup
becomes very similar to the ones in these studies.

%FIG12
\begin{figure}\begin{center}
\includegraphics[width=\columnwidth]{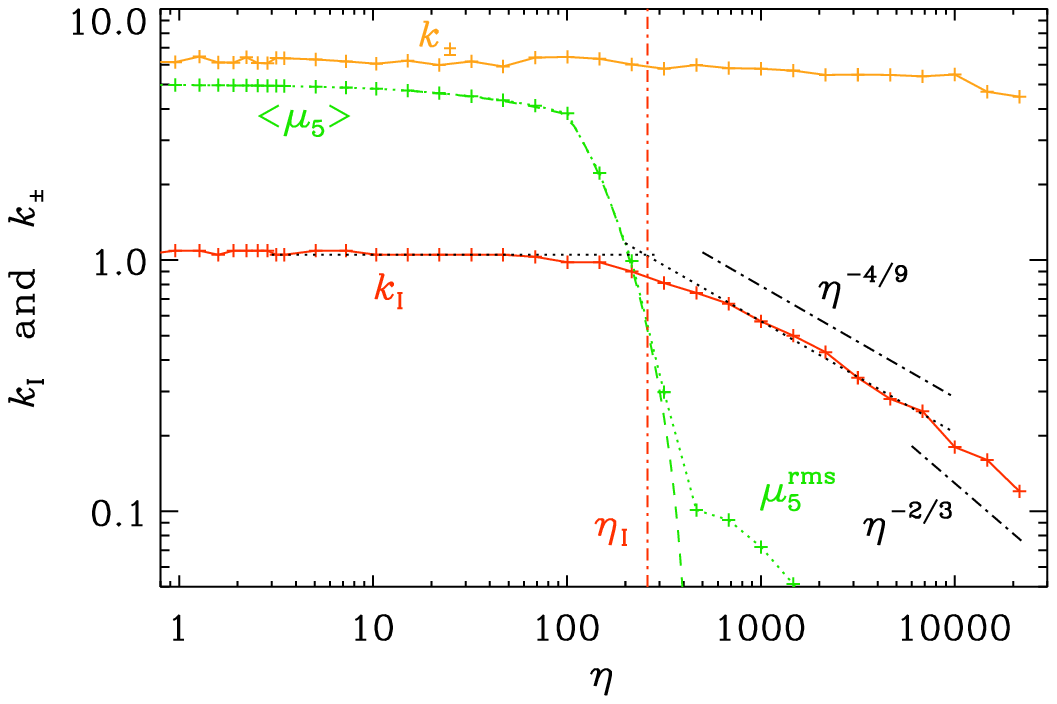}
\end{center}\caption[]{
$k_\pm^\mathrm{(i)}$ and $k_{\rm I}$ for Run~F with spin flipping,
$\eta_{\rm flip}=100$ and $\Gamma_{\rm f0}=10^{-2}$ for the rest of the run.
As in \Fig{pkmax_comp_1024a_mu10_k002o}, the green dashed line shows
$\aver{\mu_5}$ and the green dotted line shows $\mu_5^{\rm rms}$.
}\label{pkmax_comp_1024d_mu5_Gam1em2_lam2e4}\end{figure}

In \Fig{pkmax_comp_1024d_mu5_Gam1em2_lam2e4}, we show
$\eta_\pm^\mathrm{(i)}$ and $\eta_{\rm I}$ in the presence of
spin flipping.
The results suggests that the $\eta^{-4/9}$ decay changes into
the faster $\eta^{-2/3}$ decay.
Spin flipping brings $\aver{\mu_5}$ close to zero.
This process stops or slows down the decline of magnetic helicity, which
therefore remains positive.
At late times, $\aver{\mu_5}$, which was originally negative, now becomes
positive and settles at a value of around $\aver{\mu_5}\approx k_1$.
This is because at later times the positive chirality induced due
to the helicity decay by the magnetic diffusion through the chiral
anomaly is balanced by the erasure of the chirality through the
CME~\cite{Hirono+15,Schober+2020}, similar to the baryon asymmetry through
the magnetic helicity decay much before the electroweak phase transition
\cite{Giovannini:1997eg, Giovannini:1997gp, Fujita:2016igl}.

The sign of the final value of $\aver{\mu_5}$ is determined by the
magnetic helicity after the decay of $\aver{\mu_5}$ due to the onset of
spin flipping.
In the cases presented above, the sign of the magnetic helicity at the
time of the onset of spin flipping was positive and thus the chiral
chemical potential at later times was also positive.
If the initial magnetic field is weaker and the total chirality being
negative (see App.~\ref{Imbalanced}), the sign of the final value of
$\aver{\mu_5}$ can stay negative.

Our runs show that spin flipping can lead to a significant increase of the fraction
of the magnetic helicity that can be preserved in spite of the fact that
the system has vanishing total chirality.
This also reduces the total energy density dissipation of the system.
In the absence of spin flipping, both magnetic helicity and chiral chemical
potential would approach zero, so there would be no magnetic helicity available
for successful baryogenesis.
In the real Universe, however, spin flipping due to the electron Yukawa interaction,
which really violates the (total) chirality conservation, 
inevitably acts at $T\lesssim 10^2\,$TeV \cite{Campbell:1992jd,Bodeker:2019ajh}, 
and hence magnetic helicity survives more or less at the electroweak phase transition. 

In \Fig{pkmax_comp_1024d_mu5_Gam1em2_lam2e4}, we see an interval
between the onset of spin flipping, $\eta=\eta_\mathrm{flip} = 10^2$,
and the onset of the $\eta^{-2/3}$ scaling evolution of
$k_{\rm I}$, $\eta \sim 6 \times 10^3$, which
marks the real onset of the evolution with
(pure) magnetic helicity conservation.
For a rough estimate of the magnetic field evolution, however, 
we shall practically use $\eta_\mathrm{flip}$ as the 
switching time between the adapted Hosking integral conservation 
and the (pure) magnetic helicity conservation.

\subsection{Cases with initially small $|\mu_{50}|/k_0$} \label{sec:IIIH}

In all the cases considered so far, we assumed $|\mu_{50}|/k_0 > 1$.
We now consider the opposite case and discuss runs with $\mu_{50}=-0.1$,
keeping still $k_0=1$, so $|\mu_{50}|/k_0=0.1$ (Runs~P, Q, and R), and
also a run with $\mu_{50}=-0.5$ and $k_0=1$ (Run~S).
To prevent the magnetic field from being too weak, while still preserving
vanishing total chirality, we decrease the value of $\lambda$
and choose $\lambda=500$, 50, and 5 for Runs~ P (and S), Q, and R,
respectively.
All the runs end at $\eta\sim 10^4$.
The parameters of these runs are summarized in \Tab{Tbeta}.
The magnetic diffusivity is taken as $\sigma^{-1} = 2 \times 10^{-4}$. 
In all our cases, the system does not even reach $\eta_\mathrm{CPI}$.
This is because of the small value of $|\mu_{50}|$, which enters the
CPI time inverse quadratically; see Eq.~\eqref{etaCPIdiff}.

\begin{table}[t]\caption{
Empirical values of $\beta$ for cases with $|\mu_{50}|<k_0$.
For a given value of $\lambda$, the values of $\vAz$ followed
from the requirement that the total chirality vanishes.
The resulting maximum rms velocity $\urms^{\max}$ is listed for
completeness.
}\vspace{12pt}\centerline{\begin{tabular}{cccccccccc}
Run & $-\mu_{50}$ & $\lambda$ & $v_\mu$ & $\tilde{v}_\lambda$ & $\vAz$ & $\urms^{\max}$ & SLD & $\beta$ & $\eta_{\rm I}$ \\
\hline
P & 0.1 &500 & $2\times10^{-5}$ & 0.014 & 0.026 & 0.008 &  no & 0.33 & 160 \\% 1024a_mu01_k002o
Q & 0.1 & 50 & $2\times10^{-5}$ & 0.045 & 0.081 & 0.028 &  no & 0.15 &  50 \\% 1024a_mu01_k002r
R & 0.1 &  5 & $2\times10^{-5}$ & 0.141 & 0.257 & 0.076 & yes & 0.05 &  27 \\% 1024a_mu01_k002q2
S & 0.5 &500 & $       10^{-4}$ & 0.032 & 0.057 & 0.019 &  no & 0.33 &  70 \\% 1024a_mu05_k002o
\label{Tbeta}\end{tabular}}\end{table}

\begin{table}[t]\caption{
Possible combinations of $q=2/(\beta+3)$, $p=2(1-q)$, and $r=p-q$
in the range $0\leq\beta\leq1/3$.
}\vspace{12pt}\centerline{\begin{tabular}{llll}
$\qquad\beta$ & $\qquad q$ & $\qquad p$ & $\qquad r$ \\
\hline
$3/2=1.50$        & $4/9\approx0.44$   & $10/9\approx1.11$ & $2/3\approx0.67$  \\
$1/3\approx0.33$  & $3/5=0.60$         & $4/5=0.80$         & $1/5=0.20$         \\
$0.15          $  & $    0.63$         & $    0.73$         & $0.10    $         \\
$0.05          $  & $    0.66$         & $    0.69$         & $0.03    $         \\
$0             $  & $2/3=0.67$         & $2/3=0.67$         & $0       $         \\
\label{Tpqr}\end{tabular}}\end{table}

%FIG13
\begin{figure}[t!]\begin{center}
\includegraphics[width=\columnwidth]{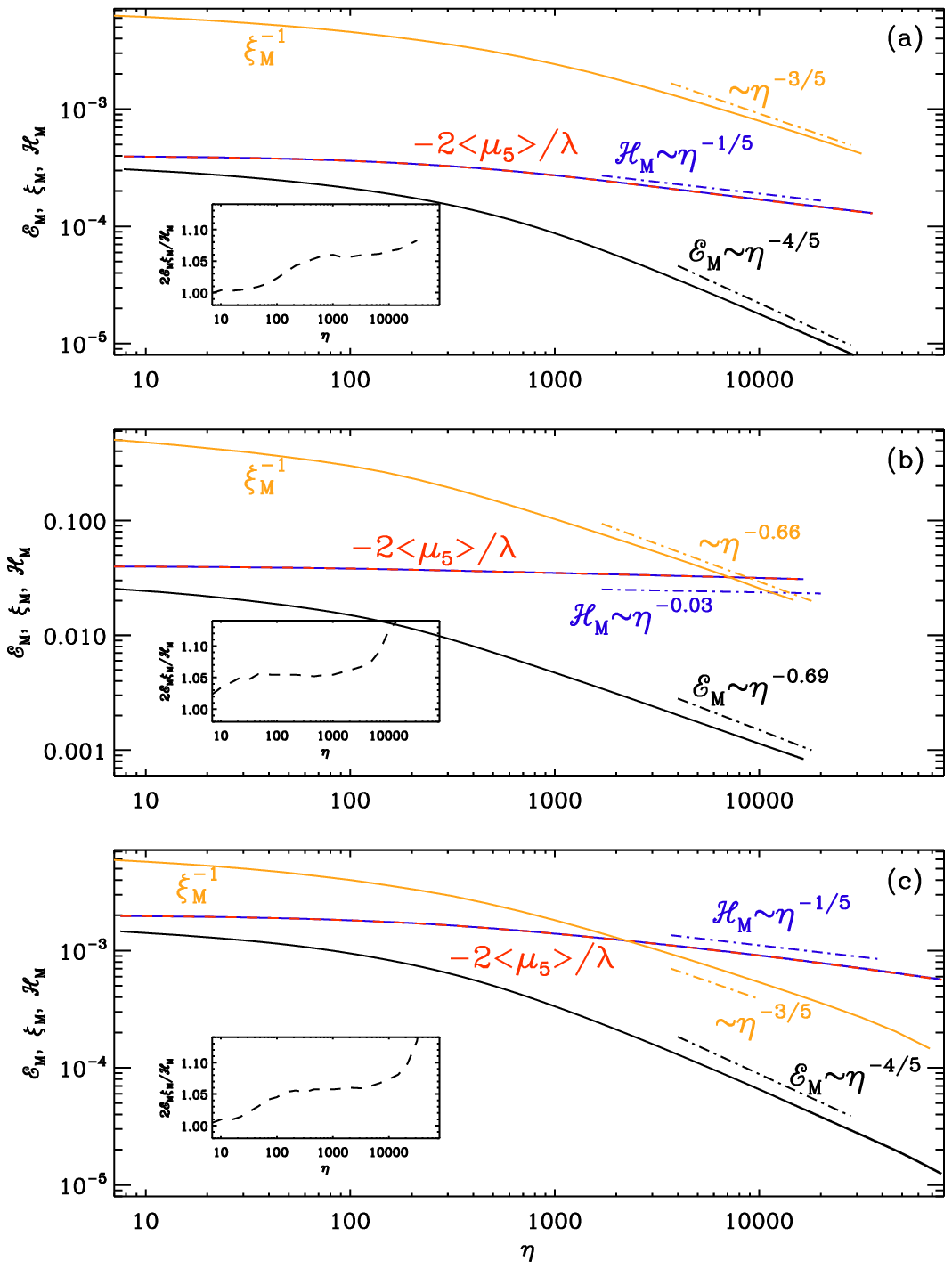}
\end{center}\caption[]{
Time dependence of $\EEM$ (black), $\xiM^{-1}$ (orange), $\HHM$ (red),
and $-2\aver{\mu_5}/\lambda$ (blue), for Runs~P (a), R (b), and S (c).
}\label{penerg_RunsPRS}\end{figure}

%FIG14
\begin{figure}[t]\begin{center}
\includegraphics[width=\columnwidth]{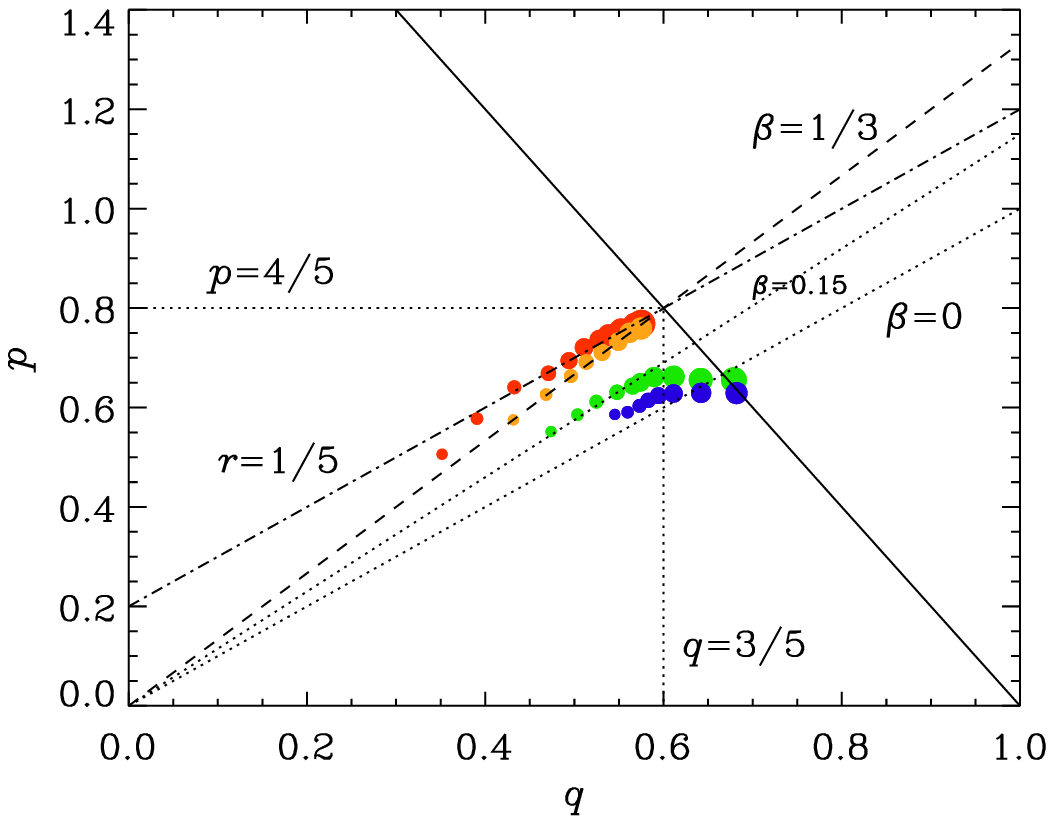}
\end{center}\caption[]{
$pq$ diagram for Runs~P (red symbols), S (orange symbols),
Q (green symbols), and R (blue symbols) at times $t=700$,
1000, 1500, 2200, 3200, 4600, 6800,
$10^4$, $1.5\times10^4$, $1.5\times10^4$, $2.2\times10^4$, and
$3.2\times10^4$, corresponding to symbols of increasing size.
The solid line denotes the scale-invariance line $p=2(1-q)$, the dashed
line the empirical $\beta=1/3$ line, and the dashed-dotted
line is the resulting $r=1/5$ line for the magnetic helicity decay.
For stronger magnetic field strength (Runs~Q and R), the solutions
evolve along $\beta\approx0.15$ and $\beta\approx0.05$, respectively.
Toward the end of the runs, the finite size effects of the domain
begin to affect the solution.
The dotted line denotes the $\beta=0$ line for magnetic helicity
conservation and is shown for comparison.
}\label{pEMxi_pq_runPQR}\end{figure}

%FIG15
\begin{figure}\begin{center}
\includegraphics[width=\columnwidth]{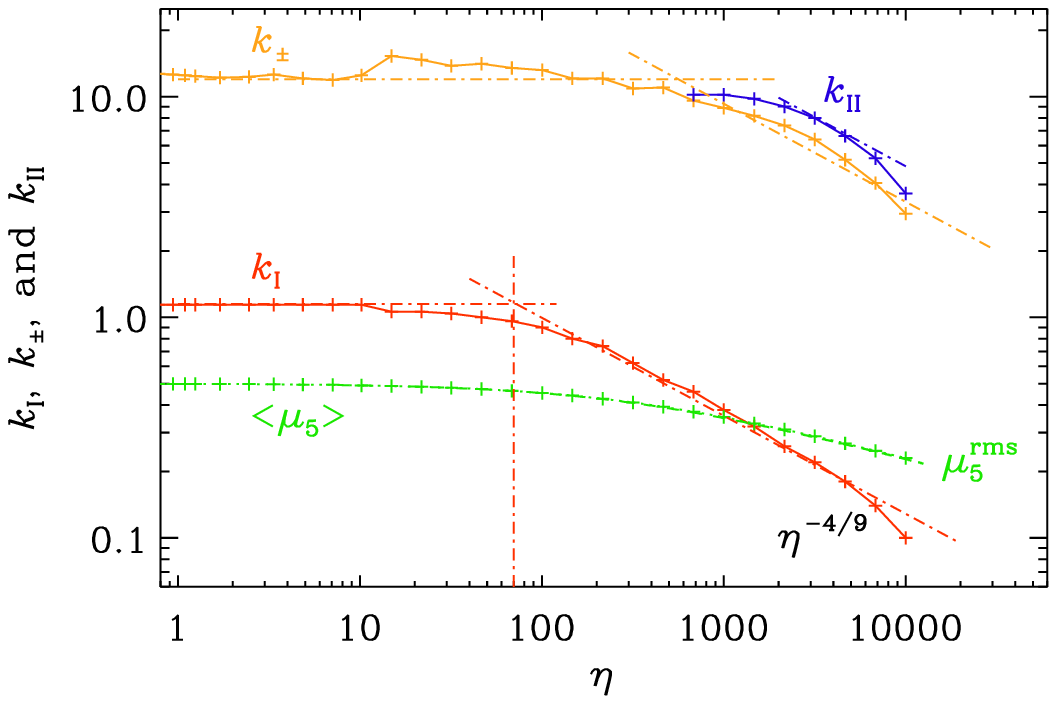}
\end{center}\caption[]{
Comparison of $k_{\rm I}$ (red), $k_\pm$ (orange),
and $k_{\rm II}$ (blue) for Run~S with $\mu_{50}=-0.5$.
The dashed-dotted line indicates $\eta^{-4/9}$.
The green dashed line shows $\aver{\mu_5}$ and the
green dotted line shows the rms value $\mu_5^{\rm rms}$.
}\label{pkmax_comp_1024a_mu05_k002o}\end{figure}

Smaller values of $\lambda$ correspond to larger magnetic fields.
We see that this also leads to a gradual decrease of 
the scaling index of the envelope of the magnetic energy spectrum,
$\beta$; see \Fig{rspec_select_1024a_mu10_k002o}.
For a given value of $\beta$, we expect that the scaling indices 
$q$, $p$, and $r$, for the
evolution of the magnetic coherence length ($\xiM\propto\eta^q$),
energy density ($\EEM\propto\eta^{-p}$), and
helicity ($\HHM\propto\eta^{-r}$), respectively,
are given as $q=2/(\beta+3)$, $p=2(1-q)$, and $r=p-q$. 
In \Fig{penerg_RunsPRS}(a), we see that for Run~P the exponents in
agree reasonably well with those expected for $\beta=1/3$.
In \Fig{penerg_RunsPRS}(b), we also show the results for Run~R, where
$\lambda$ is a hundred times smaller and the magnetic field ten times
stronger.
Now the value of $\beta$ is very small (about 0.05), corresponding
to $q=0.66$, $p=0.69$, and $r=0.03$.
Finally, \Fig{penerg_RunsPRS}(c) gives the results for Run~S, where
$\lambda=500$ is the same as for \Fig{penerg_RunsPRS}(a), but now
$\mu_{50}=-0.5$ instead of $-0.1$.
In this case, similarly to Run~P, $\beta\approx1/3$,
$p\approx4/5$, $q\approx3/5$, and $r\approx1/5$.

In \Tab{Tpqr}, we list several combinations of the expected scaling
indices $q$, $p$, and $r$ for $0\leq \beta \leq 3/2$.
Interestingly, in the range $0\leq\beta\leq1/3$, the values of $q$
and $p$ do not vary much in this range, especially compared to the case 
for the evolution with the (adapted) Hosking integral conservation, $\beta=3/2$,
so if they do not agree with
those from the simulations, the discrepancy cannot easily be resolved
by changing the value of $\beta$ within reasonable limits.
Note that $\beta=0$ is expected if the evolution is governed by
(pure) magnetic helicity conservation.

In the corresponding $pq$ diagram \Fig{pEMxi_pq_runPQR}, we see that
all the runs approach the scale-invariance line $p=2(1-q)$.
For Run~P, it evolves along the line $r=1/5$.
At the intersection, we have $q=3/5$ and $p=4/5$.
However, for Runs~Q and R with stronger magnetic field strengths,
expressed in terms of the initial Alfv\'en speed $\vAz=B_0/\sqrt{\rho_0}$
(which is well approximated by $\tilde{v}_\lambda$), the solution
approaches the $\beta=0$ line, which suggests better conservation of
magnetic helicity.
Note that near the end of those runs, the data points may not be
reliable because of the finite size of the domain.
In addition, because of the stronger magnetic field, the Alfv\'en time
is shorter and therefore $k_{\rm I}$ reaches $k_1$ more quickly.
In any case, it is likely that for small $|\mu_{50}|/k_0$, we see an
new stage of the evolution of the system where the magnetic
helicity and chirality are temporarily conserved individually, as
discussed in Sec.~\ref{IIE}.
This is supported by the fact that the theoretically predicted time
of the onset of ACC (or even CPI) should come much later, well after
the end of the run; see \Eqs{etaCPIdiff}{etaacck0>mu50}.
In other words, from the present simulation results, we cannot distinguish
between the two possible theoretical predictions for the later
evolution in the case $k_0 > |\mu_{50}|$; see steps~3 or $3^\prime$
given in Sec.~\ref{IIE}.

For all the simulations where initially $|\mu_{50}|<k_0$,
we find that $\mu_5$ decays more slowly than $k_{\rm I}$; see
\Fig{pkmax_comp_1024a_mu05_k002o} for Run~S, as an example,
where we see the crossing of $\mu_5$ and $k_{\rm I}$.
The same is also seen for $|\mu_{50}|=0.1$, but then the crossing
of $|\mu_5|$ and $k_{\rm I}$ occurs much later and is therefore not
yet prominent.
Again, these observations suggest that the magnetic
helicity-conserving phase is an intermediate one before the solution
resumes the decay governed by the adapted Hosking integral, as discussed
in Sec.~\ref{IIE}.

%FIG16
\begin{figure}[t!]\begin{center}
\includegraphics[width=\columnwidth]{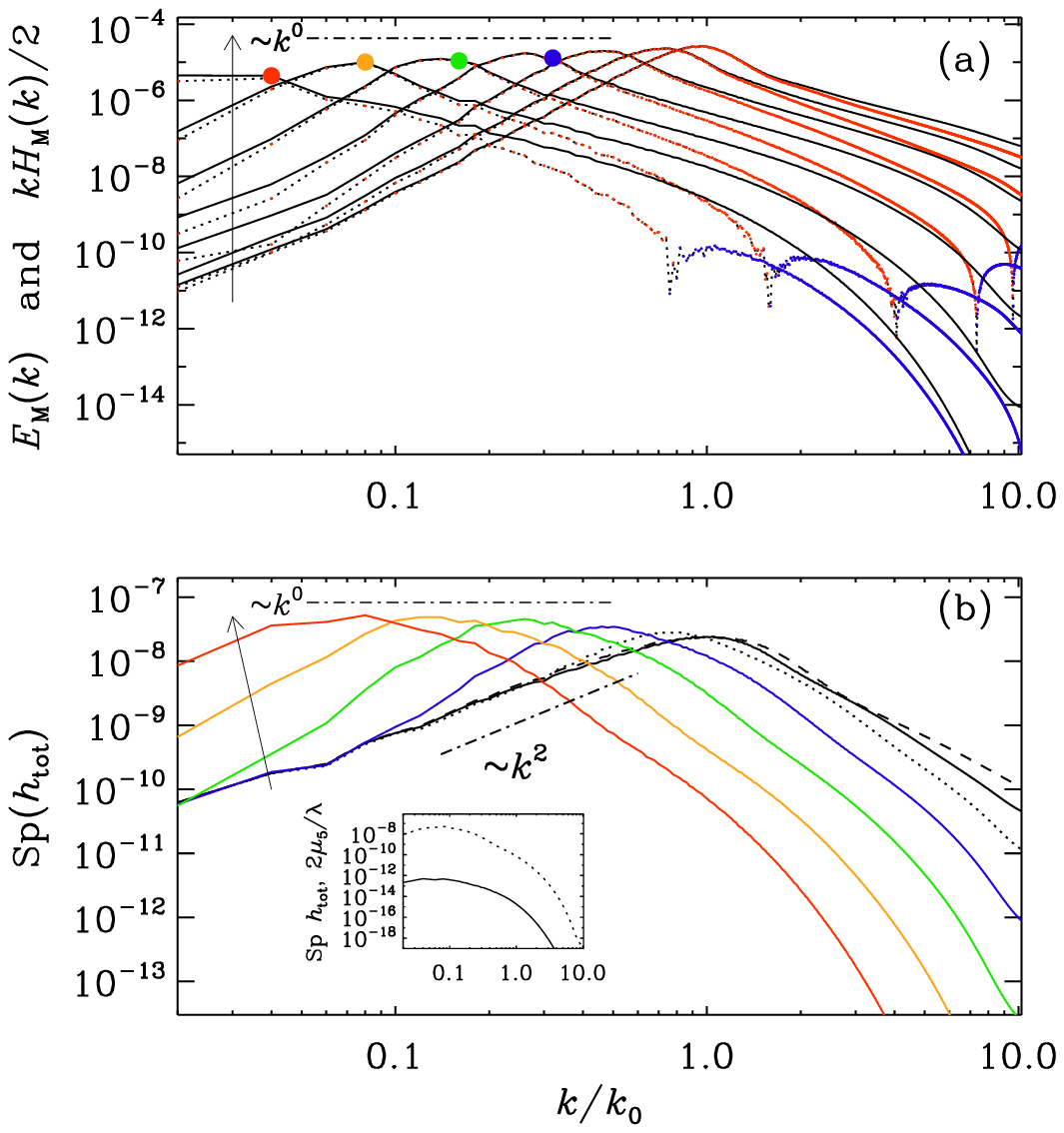}
\end{center}\caption[]{
(a) Magnetic energy and (b) total helicity variance spectra at $t=70$ (dashed),
$200$ (solid), $700$ (dotted), $2\times10^3$ (blue), $7\times10^3$ (green),
$2\times10^4$ (orange), and $7\times10^4$ (red) for Run~S.
In (a), note that the $\EM(k,t)$ evolve underneath a $k^{1/3}$ envelope,
and the upward arrow indicates the sense of time.
In (b), the slopes $k^2$, $k^4$, and $k^{-4}$ have been indicated
and the inset compares $\Sp(2\mu_5/\lambda)$ (solid) with
$\Sp(h_{\rm tot})$ (dotted) at the last time.
}\label{rspec_select_HoskM_1024a_mu01_k002o}\end{figure}

%FIG17
\begin{figure}[t!]\begin{center}
\includegraphics[width=\columnwidth]{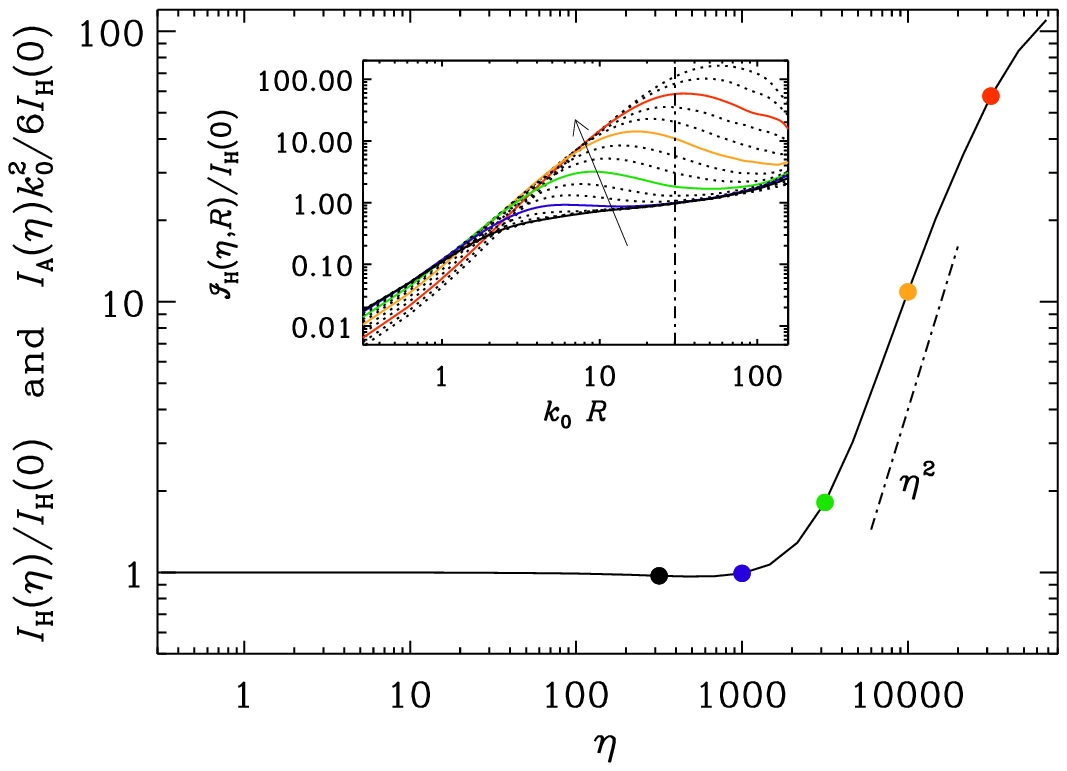} %(Run S)
\end{center}\caption[]{
$I_{\rm H}(t)$, normalized by its initial value, for Run~P.
The inset shows ${\cal I}_{\rm H}(R,t)$ versus $R$ at different times:
solid lines correspond to $\eta=320$, 1000, 3200, 10,000, and 32,000
are in black, blue, green, orange, and red, respectively.
The dotted lines mark intermediate times.
The colors of the symbols in the graph of $I_{\rm H}(t)$
correspond to those of the lines in the inset, and also to
those in \Fig{rspec_select_HoskM_1024a_mu01_k002o}.
The adapted Hosking integral is evaluated as $I_{\rm H}(t)={\cal I}_{\rm H}(R_*,t)$.
The vertical dashed-dotted line marks the value $k_0 R_*=100$, which is where the
curves show a plateau at early times.
}\label{psaff_1024a_mu01_k002o_rr}\end{figure}

The time evolution of the magnetic energy and helicity spectra for Run~P
are given in \Fig{rspec_select_HoskM_1024a_mu01_k002o}(a).
We can see that a negative magnetic helicity part of the spectrum still
emerges, again only at large wave numbers, although now much later.
This means that $|\aver{\muM^-}|$ is induced by the CPI, but it stays
extremely small.
However, the time is still less than $\eta_\mathrm{CPI}$ 
and hence it is likely that we only see the early stages of the CPI
before much stronger amplification is possible.
Furthermore, $|\aver{\muM^+}|$ does not decay much during the time of the
run; see also \Fig{penerg_RunsPRS}(b).
This can easily be understood by the fact that
$\eta_{\rm CPI}$ is very large in this run.
Many other features of the magnetic field evolution remain superficially
similar to the limit of large $|\mu_{50}|/k_0$.
One still sees inverse cascading of positive magnetic helicity.

\subsection{Non-conservation of $I_{\rm H}(\eta)$ for small $|\mu_{50}|/k_0$}

In \Fig{rspec_select_HoskM_1024a_mu01_k002o}(b), we show 
the total helicity variance spectrum $\Sp(h_{\rm tot})$.
We clearly see that for small $k$, the value of $\Sp(h_{\rm tot})$ 
increases with time.
This suggests that the adapted Hosking integral, as defined in
Ref.~\cite{BKS23}, is not conserved at late times.

In \Fig{psaff_1024a_mu01_k002o_rr}, we show for Run~S
the Hosking integral, $I_{\rm H}$, as a function of time.
It is obtained from the function ${\cal I}_{\rm H}(R,\eta)=
L^{-3}\int w(\kk,R)\,\Sp(h_{\rm tot})\,\dd^3 \kk/(2\pi)^3$, which is shown in
the inset as a function of $R$ at different times.
Here, $w(k,R)=(4\pi R^3/3)[6j_1(kR)/kR]^2$ is a weight function
\citep{Zhou+22} with $j_n$ being spherical Bessel functions.
The relevant value of $R$ is usually where $R$ is just a little less
than half the system size \cite{Hosking+Schekochihin21, Zhou+22},
which is here for $k_0 R=50\pi\approx157$.
At that location, ${\cal I}_{\rm H}(R,\eta)$ usually also shows an
approximate plateau for different times.
This is not the case for this run.
We have therefore chosen instead the value $k_0 R=30$, which is where
${\cal I}_{\rm H}(R,\eta)$ shows considerable spread in time.
This position is marked in the inset by the vertical dashed line.
Focussing now on the resulting time dependence of $I_{\rm H}(\eta)$, we see a sharp
rise at late times ($\eta\geq10^4$), which also agrees with the time
when we saw the values of $\Sp(h_{\rm tot})$ for small $k$ to change;
see \Fig{rspec_select_HoskM_1024a_mu01_k002o}(b).

Regarding the conserved quantity for runs in the limit of small
$|\mu_{50}|/k_0$, we can say that, in spite of vanishing total chirality,
the Hosking integral is here not conserved, because the magnetic energy
now peaks at scales where the CME is not effective during the time of
the run and the magnetic helicity is conserved.
As a result, the net chirality is no longer random, but systematically
of positive sign.
The subinertial range of the magnetic helicity variance begins to be
dominated by a $k^4$ spectrum, which suggests that the Hosking integral
in the expansion $\Sp(h_{\rm tot})=I_{\rm H} k^2/2\pi^2+O(k^4)$ is now
subdominant.

To summarize, these runs are consistent with the theoretical prediction
in Sec.~\ref{IIE}, up to the intermediate stage (step~2), although a moderate violation of helicity
conservation has been seen for Run~P.
Note that Run~P has a larger value of $\lambda$, which makes
the theoretically predicted $\eta_\mathrm{ACC}$ smaller [see
\Eq{etaacck0>mu50}], so that an earlier transition to the evolution
with adapted Hosking integral conservation is expected.
For an analytic estimate of the evolution of the system in the next section, 
we shall use the theoretical prediction discussed in Sec.~\ref{IIE}. 
Namely, the system is frozen until $\eta=\eta_\lambda$ and then evolves
with the usual inverse cascade for the helical magnetic field as an
intermediate stage. 
We chose step~3$^\prime$ for the onset of the ACC, 
which is more realistic, though at present we do not have 
any support from numerical results.
That is, at $\eta=\eta_{5\mathrm{dec}}$ the system
starts to evolve with a decay law determined by the conservation
of the adapted Hosking integral.

%%%%%%%%%%%%%%%%%%%%%%%%%%%%%%%%%%%%%%%%%%%%%%%%%%%%%%%%%%%%%%%%%%%%%%%%%%%%%%%%%%%%%%%%%%%%%%%%%%%%

\section{Application to the early Universe}
\label{Application}

%%%%%%%%%%%%%%%%%%%%%%%%%%%%%%%%%%%%%%%%%%%%%%%%%%%%%%%%%%%%%%%%%%%%%%%%%%%%%%%%%%%%%%%%%%%%%%%%%%%%

\subsection{From QED to the Standard Model}

Now we investigate the impact of our findings in the previous sections
on the cosmology of the early Universe, especially, baryogenesis.
Up to here, we focused on a QED-like theory.
Thus, we first would like to clarify
its relation to the dynamics in the early Universe. The SM
involves the right-handed
leptons $e_{Rf}$, the left-handed lepton doublets $\ell_{Lf}$, the
right-handed up- and down-type quarks, $u_{Rf}$ and $d_{Rf}$, and the
left-handed quark doublets $q_{Lf}$ with the flavor index running through
$f=1,2,3$, alongside the scalar Higgs doublet $\Phi$, which are in total
$16$ species.
On top of this, we have gauge interactions of
$\mathrm{U}(1)_Y \times \mathrm{SU}(2)_L \times \mathrm{SU}(3)_C$.
It is not obvious why we can reduce this complicated system to chiral
MHD based on a QED-like theory like the one introduced in Sec.~\ref{ChiralMHD}.

What we are interested in here is the slow dynamics at long wave lengths
compared to interactions among particles.
The key idea for the reduction is to assume the equilibration of fast
interactions and to keep only the slow variables.
The hypermagnetic field of $\mathrm{U}(1)_Y$ with a correlation length
much larger than the mean free path of the particles stands out as a slow
variable because the magnetic flux cannot be cut thanks to the absence
of monopoles.
This feature does not hold for non-Abelian gauge fields because they are
charged under their own gauge group.
We also need the chiral chemical potential, since it is related to the
magnetic field via the anomaly equation.
Apart from these two fundamental building blocks, we can coarse-grain
the microscopic properties of all particle in the form of transport
coefficients such as the diffusion constant and the electric conductivity,
besides macroscopic quantities such as the pressure, energy density,
and velocity field. In this way, one may see that the system can be reduced to chiral MHD
as far as the slow and long-wave dynamics is concerned.

Still, one might wonder why we can just focus on one particular chiral
chemical potential, as in Eq.~\eqref{dAdt}, since we have $15$ chiral fermion species in the SM.
To illustrate this, let us focus on the temperature right above $10^5\,
\mathrm{GeV}$, where the electron Yukawa interaction is not efficient
compared to the cosmic expansion, but other interactions are fast enough.
In this case, the chiral chemical potential for the right-handed electron,
$\tilde{\mu}_{e}$, should be counted as a slow variable, as it is directly
related to the hypermagnetic field via the anomaly equation.
On the other hand, other chiral chemical potentials are subject to fast
SM interactions, which provides $11$ nontrivial constraints among them.
Recalling that the SM has four conserved charges, hypercharge $Y$ and
the flavored baryon-minus-lepton numbers $B/3 - L_f$ with $f = 1,2,3$,
one may immediately see that the remaining $15$ chemical potentials
can be expressed as a function of $\tilde{\mu}_e$ by solving $11 +
4$ constraints.
The chiral chemical potential $\tilde{\mu}_5$
originates from the generalized Ohm's law,
\begin{equation}
\bm{J}_Y = \sigma_Y \bm{E}_Y + \frac{2 \alpha_Y}{\pi}\tilde{\mu}_5\,\bm{B}_Y,
\end{equation}
where $\alpha_Y$ is now the $\mathrm{U}(1)_Y$ fine-structure constant
and $\sigma_Y$ the hyperelectric conductivity of the plasma.
In the following, we will work with the $\alpha_Y$ value around the electroweak
scale, $\alpha_Y \simeq 0.01$, and neglect its renormalization group running when
considering the dynamics of the hypermagnetic field at high energies.
Also, note that, in this section, we set $\hbar = c = \kB = 1$, and all
quantities are physical rather than comoving, unless explicitly stated otherwise.
For the SM $\mathrm{U}(1)_Y$, at $T \sim 10^{5\cdots6} \mathrm{GeV}$, one may express this $\tilde{\mu}_5$ as a summation
of chiral chemical potentials for the SM fermions as \cite{JS97,
Domcke:2022kfs}
\begin{equation}
\label{eq:mu5Y}
\tilde{\mu}_5 = \sum_{i=1}^{15} \epsilon_i\,g_i\,y_i^2\,\frac{\tilde{\mu}_i}{2} = \frac{711}{481}\,\frac{\tilde{\mu}_e}{2} \,,
\end{equation}
where $i$ runs over all SM fermions, $\epsilon_i = \pm$ for right-
and left-handed fermions, respectively, $g_i$ counts internal degrees
of freedom, and $y_i$ is the hypercharge of fermion species $i$. In
the second equality, we inserted the solution of the $15$ constraint
equations mentioned above. We now see that, up to the $\mathcal{O}(1)$
coefficient of $711/481$, one chiral chemical potential suffices to
describe the system.
In higher temperature regimes, we will have additionally more slow
variables that enter the expression of $\tilde{\mu}_5$, but it is
still written as a linear combination of their chemical potentials with
$\mathcal{O}(1)$ coefficients.
It still holds that the evolution of the system is described by chiral
MHD as discussed in Sec.~\ref{ChiralMHD} with $\tilde{\mu}_5$ being
evaluated accordingly.

 %%%%%%%%%%%%%%%%%%%%%%%%%%%%%%%%%%%%%%%%%%%%%%%%%%%%%%%%%%%%%%%%%%%%%%%%%%%%%%%%%%%%%%%%%%%%%%%%%%%%

\subsection{Baryogenesis}
\label{Baryogenesis}

After these general remarks, let us now turn to the implications of our
analysis for the generation of the baryon asymmetry of the Universe.
We are primarily interested in the scenario of baryogenesis from decaying
hypermagnetic helicity~\cite{Giovannini:1997eg,Giovannini:1997gp,Fujita:2016igl,Kamada:2016cnb,Kamada:2016eeb},
which assumes the presence of a strongly helical hypermagnetic field
during the radiation-dominated era in the early Universe.
This scenario is based on the observation that the helicity stored in the
hypermagnetic field decays at the time of the electroweak phase transition,
not because of some exotic helicity-violating interactions, but simply
because hypermagnetic helicity is converted to magnetic helicity.
This decay of hypermagnetic helicity then sources a baryon asymmetry
via the chiral anomaly of the baryon-number current.

One possibility to generate the helical hypermagnetic field required for baryogenesis consists of axion inflation featuring a Chern--Simons coupling to $\mathrm{U}(1)_Y$.
Such a model leads to the nonperturbative production of hypermagnetic gauge fields in combination with charge asymmetries for the 15 chiral SM fermion species~\cite{Domcke+19, Domcke:2022kfs},
\begin{equation}
\label{eq:ninibar}
n_i - \bar{n}_i = \frac{1}{6}\,g_i\,\tilde{\mu}_i\,T^2 = - \epsilon_i\,g_i\,y_i^2\,\frac{\alpha_Y}{2\pi}h_Y + \cdots \,,
\end{equation}
where the ellipsis represents all other SM contributions, which, however,
can safely be neglected during inflation.%
\footnote{The top-quark Yukawa interaction would be a
possible exception; see the discussion in footnote~$5$
of Ref.~\cite{Domcke:2022kfs} for more details.}
Furthermore, $h_Y$ in Eq.~\eqref{eq:ninibar} is the physical helicity
density, which we define in terms of the comoving vector potential
$\bm{A}_{Y,\rm com}$, comoving hypermagnetic field $\bm{B}_{Y,\rm com}$,
and scale factor $a$,
\begin{equation}
h_Y = \frac{1}{a^3}\, \langle \bm{A}_{Y,\rm com} \cdot \bm{B}_{Y,\rm com} \rangle \,, 
\end{equation}
where the angle brackets now stand for a double average including the spatial average and the quantum mechanical expectation value during inflation. From Eq.~\eqref{eq:ninibar}, we can read off the fermion chemical potentials at the end of inflation in terms of the helicity density at the end of inflation.
Inserting this result into Eq.~\eqref{eq:mu5Y}, we obtain the chiral
chemical potential at the end of inflation,
\begin{equation}
\label{eq:mu5ini}
\frac{\tilde{\mu}_5}{T} = -\frac{c_5}{2}\,6\,\chi \,, \qquad c_5 = \frac{95}{18} \,,
\end{equation}
where the dimensionless yield parameter $\chi$ quantifies the amount of $CP$ violation during axion inflation~\cite{Domcke:2022kfs},
\begin{equation}
\label{eq:chi}
\chi = \frac{\alpha_Y}{2\pi}\frac{h_Y}{T^3} \,.
\end{equation}
Here, we assume instantaneous reheating.
The same coefficient $c_5$ was found in Ref.~\cite{Domcke:2022kfs}; in
total, the expression for $\tilde{\mu}_5$ in Eq.~\eqref{eq:mu5ini} is,
however, smaller than the one in Ref.~\cite{Domcke:2022kfs} by a factor
of $\sfrac{1}{2}$ because, in the present paper, we include a factor of
$\sfrac{1}{2}$ in Eq.~\eqref{eq:mu5Y}.

The fermion asymmetries generated during axion inflation are consistent with the chiral anomalies of the respective fermion currents. In fact, it is straightforward to generalize the conversion law in Eq.~\eqref{eq:mutot} to the early Universe. To see this, let us rewrite Eq.~\eqref{eq:mu5ini} as follows,
\begin{equation}
\mu_5 + \frac{3\,c_5}{2} \left(\frac{2\alpha_Y}{\pi}\right)^2\frac{1}{2a^3T^2}\, \langle \bm{A}_{Y,\rm com} \cdot \bm{B}_{Y,\rm com} \rangle = 0 \,,
\end{equation}
where we used $\mu_5 = \left(2\alpha_Y/\pi\right)\tilde{\mu}_5$. Then, introducing
\begin{equation}
\lambda_Y = 3\left(\frac{2\alpha_Y}{\pi aT}\right)^2 \,,
\end{equation}
we obtain the relation
\begin{equation} \label{conseqSM}
\mu_5 + \frac{c_5}{2}\,\mu_{\rm M}^Y = 0 \,,
\end{equation}
where 
\begin{equation}
\mu_{\rm M}^Y = \frac{1}{2a}\lambda_Y \langle \bm{A}_{Y,\rm com} \cdot \bm{B}_{Y,\rm com} \rangle \,.
\end{equation}

As the temperature in the early Universe decreases, more and more
SM interactions reach chemical equilibrium.
This includes the SM Yukawa interactions, which violate parity and hence
render the coefficient $c_5$ in Eq.~\eqref{eq:mu5ini} a time-dependent
quantity~\cite{Domcke:2022kfs}.
During axion inflation, $c_5$ assumes its maximal value,
$c_5 = 95/18 \simeq 5.3$, before it then decreases down to
$c_5 = 711/481 \simeq 1.5$ at temperatures of a few 100 TeV
[see Eq.~\eqref{eq:mu5Y}].
This change in $c_5$ is reflected in a changing value of the
chiral chemical potential $\mu_5$, which is always given by
$\mu_5 = -c_5/2\,\mu_{\rm M}^Y$ according to Eq.~\eqref{conseqSM},
with $\mu_{\rm M}^Y$ remaining constant until the onset of CPI, ACC,
or electroweak phase transition.
At $T \lesssim 10^5\,\textrm{GeV}$, $c_5$ and hence $\mu_5$ vanish
because all SM interactions have reached chemical equilibrium.

The $CP$ asymmetry parameter $\chi$ in Eq.~\eqref{eq:chi} controls the
outcome of baryogenesis from helicity decay.
That is, if no CPI or ACC takes place before the onset of spin flipping,
the decay of hypermagnetic helicity around the electroweak phase transition
results in a present-day baryon asymmetry (quantified in terms of the baryon-to-photon ratio) that is fully controlled by
$\chi$ \cite{Domcke:2022kfs},
\begin{equation}
\label{eq:etaold}
\eta_B^0 \equiv \frac{n_B^0}{n_\gamma^0} \simeq 0.15\,c_B^{\rm dec} \chi \,,
\end{equation}
where $n_\gamma = 2\,\zeta(3)T^3/\pi^2$ and the superscript $0$ indicates that 
a quantity is evaluated at the present time.
Here, the coefficient $c_B^{\rm dec}$ has a theoretical uncertainty of possibly two orders of magnitude~\cite{Kamada:2016cnb}.
In the following, we will work with the representative value
$c_B^{\rm dec} = 0.05$~\cite{Kamada:2020bmb,Domcke:2022kfs}, which implies that $\chi$ values of the order
of $\chi \sim 10^{-7}$ are necessary to reproduce the observed
baryon asymmetry, $\eta_B^{\rm obs} \simeq 6.1 \times 10^{-10}$~\cite{Planck:2018vyg,ParticleDataGroup:2022pth}.
Meanwhile, the parameter $\chi$ also allows us to evaluate the ratio of $k_0$ and $\mu_5$ at the end of axion inflation.
Specifically, if we estimate the comoving peak wave number $k_0$ in terms of the comoving wave number
that enters the Hubble horizon at the end of reheating,
$k_{\rm rh} = a_{\rm rh} H_{\rm rh}$~\cite{Domcke:2022kfs}, we find
\begin{align}
\label{eq:mu5krh}
\frac{\left|\mu_5\right|}{k_{\rm rh}/a_{\rm rh}} & = \frac{6\,\alpha_Y c_5\,\chi}{\pi}\frac{T_{\rm rh}}{H_{\rm rh}} = \frac{6\,\alpha_Y c_5\,\chi}{\pi}\frac{M_*}{T_{\rm rh}} \notag \\
& \sim 10^{-4}\:\bigg(\frac{\chi}{10^{-7}}\bigg) \left(\frac{10^{14} \mathrm{GeV}}{T_{\rm rh}}\right) \,,
\end{align}
where $M_* = \left(90/\pi^2/g_*\right)^{1/2}M_{\rm Pl} \simeq 7.1 \times 10^{17}\,\textrm{GeV}$
is the reduced Planck mass, $M_{\rm Pl} \simeq 2.4 \times 10^{18}\,\textrm{GeV}$,
rescaled by the effective number of relativistic degrees of freedom in
the Standard Model plasma, $g_* = 427/4$.
Axion inflation typically results in small values of the $\chi$
parameter (e.g., $\chi \sim 10^{-7}$; see above) and large values of the
reheating temperature (e.g., $T_{\rm rh} \sim 10^{14}\,\textrm{GeV}$;
see Ref.~\cite{Domcke:2022kfs}), which puts us in the parametric regime
where $\left|\mu_5\right| \ll k_{\rm rh}/a_{\rm rh}$ at the end of
axion inflation.
Moreover, smaller values of $T_{\rm rh}$ typically result in
smaller values of $\chi$, following the scaling relation $\chi \propto (T_{\rm rh}/M_*)^3$~\cite{Domcke:2022kfs}, which means that the opposite hierarchy,
$\left|\mu_5\right| \gg k_{\rm rh}/a_{\rm rh}$, cannot simply be obtained
by considering a smaller reheating temperature.

For $\left|\mu_5\right| \ll k_{\rm rh}/a_{\rm rh}$, the chiral chemical
potential eventually becomes larger than the peak wave number of the
hypermagnetic energy spectrum.
This is because the peak momentum decreases via the inverse cascade, while the chiral chemical potential is approximately conserved.
As mentioned above Eq.~\eqref{eta5dec}, we expect a large hierarchy between $\mu_5$ and $k_\text{I}$ at $\eta_\text{CPI}$ for the SM plasma,
\begin{align}
    \left. \frac{|\mu_5|}{k_\text{I} / a} \right|_\text{CPI}
    &\simeq 
    \left( \frac{2 \sigma_Y^2}{\bar \rho \lambda_Y} \right)^{1/3} \\
    &\sim 2 \times 10^2 
    \left( \frac{106.75}{g_\ast} \right)^{1/3}
    \left( \frac{0.01}{\alpha_Y} \right)^{2/3}\,,
\end{align}
where we evaluate the hyperelectric conductivity $\sigma_Y$
as $\sigma_Y = c_{\sigma_Y} T$, with $c_{\sigma_Y} \sim
10^2$~\cite{Baym:1997gq,Arnold:2000dr}, the average radiation energy
density $\bar{\rho}$ as $\bar{\rho} = c_{\bar{\rho}}\,T^4$, with
$c_{\bar{\rho}} = \pi^2 g_*/30$, and the parameter $\lambda_Y$
as $\lambda_Y = c_{\lambda_Y}/(aT)^2$ with
$c_{\lambda_Y} = 12\,\alpha_Y^2/\pi^2$.
The net chiral chemical potential may start to decay after some duration of CPI as given in Eq.~\eqref{eta5dec}
\begin{equation} \label{eta5dec-SM}
    \eta_{5\mathrm{dec}} = c_A \,\eta_\mathrm{CPI}
    = c_A \left[ \frac{1}{a} \frac{\sigma_Y}{\mu_5^2} \right]_\text{rh}\,,
\end{equation}
where the factor of $a^{-1}$ follows from the mass dimension of the factor $\sigma_Y/\mu_5^2$.

The time $\eta_{5\mathrm{dec}}$ marks the onset of the net
chirality decay via the CPI and needs to be compared to the time
$\eta_{\rm sf} = 1/k_{\rm sf}$ when spin flipping for left- and
right-handed electrons becomes efficient, where
$k_{\rm sf} = a_{\rm sf} H_{\rm sf}$ is the comoving horizon scale
at $\eta = \eta_{\rm sf}$.
Using Eq.~\eqref{eta5dec-SM}, together with
$T_{\rm sf} = 10^5\,\textrm{GeV}$ for the electron Yukawa interaction
in the SM~\cite{Campbell:1992jd,Bodeker:2019ajh}, we obtain
\begin{equation}
\label{eq:eta5decsf}
\frac{\eta_{5\mathrm{dec}}}{\eta_{\rm sf}} 
    \sim
    \bigg( \frac{g_\ast}{106.75} \bigg)^{1/2}
    \left( \frac{0.01}{\alpha} \right)^2
    \bigg(\frac{c_A}{10}\bigg) 
    \left(\frac{10^{-4}}{\chi}\right)^2 
\,.
\end{equation}
If chirality decay occurs before the onset of spin flipping, i.e., $\chi \gtrsim 10^{-4}$, we expect the hypermagnetic helicity to decay simultaneously.
Therefore, the correct amount of baryon asymmetry would be obtained for such a large $\chi \gg 10^{-7}$ that would overproduce baryon~\eqref{eq:etaold} at a first glance, since the reduction of the hypermagnetic helicity because of the CPI might counteract this overproduction of baryon number.
To see this explicitly, one may use the ratio in Eq.~\eqref{eq:eta5decsf}
to introduce a dilution factor, $\Delta = (\eta_{5\mathrm{dec}}/\eta_{\rm sf})^{q_5}$ with $q_5$ being the scaling index introduced in Eq.~\eqref{ACCscaling3},
that multiplies the naive baryon asymmetry in Eq.~\eqref{eq:etaold}
whenever the CPI should occur before the onset of spin flipping,
\begin{equation}
\label{eq:etanew}
\eta_B^0 \simeq \textrm{min}\left\{1,\Delta\right\} \times 0.15\,c_B^{\rm dec} \chi \,.
\end{equation}
By requiring $\eta_B^0 \simeq 6.1 \times 10^{-10}$, we can estimate the size of the $CP$ violation $\chi$ required to obtain the observed baryon-to-photon ratio.
For $q_5 = \sfrac{2}{3}$, we need $\chi \sim 10^6$. Such large $\chi$ values are extremely difficult, if not impossible, to realize in realistic models of axion inflation.
For $q_5 = \sfrac{4}{3}$, we need instead $\chi \sim 10^{-2}$, which is still large but not an unrealistic value for axion inflation.
In this way, the result for the final baryon asymmetry strongly depends on the scaling index $q_5$, whose precise value is, however, beyond the scope of our current simulations.

To sum up, even if we initially start from $\left|\mu_5\right| \ll k_{\rm rh}/a_{\rm rh}$, the system eventually ends up with the large hierarchy of $\left|\mu_5\right| \gg k_\text{I}/a$.
Moreover, if $\chi$ is extremely large, we can have initially $\left|\mu_5\right| \gg k_\text{I}/a$. 
These cases might display similar dynamics as our Runs~III--VI, which we commented on at the end of
Sec.~\ref{subsec:evolution}.
The decay law for the magnetic helicity may then be different from
the $\eta^{-2/3}$ behavior that we typically find for ACC, which strongly affects the outcome of baryogenesis.
As already stated in Section~\ref{subsec:evolution}, we leave a more detailed study of this more exceptional case for future work.

\section{Conclusions}
\label{Conclusions}

We have performed numerical simulations of chiral MHD 
with zero initial total chirality for a range of parameters to
determine the dependence of characteristic time and scale ratios,
which are well explained by the analytical estimate in Sec.~\ref{IIE}. 
Namely, they are consistent with the scaling evolution, $\xi_\mathrm{M} \propto \eta^{4/9}, \,
{\cal E}_\mathrm{M} \propto \eta^{-10/9}$, and $\aver{\mu_5} \propto \eta^{-2/3}$, 
derived from the conservation of the adapted Hosking integral~\cite{BKS23}, 
and also the time scale of the onset of this scaling evolution, $\eta_\mathrm{ACC}$; see Eqs.~\eqref{etaACCcase1} and \eqref{etaacck0>mu50}.
Our numerical simulations also assess the possibility of artifacts resulting from insufficient
scale separation.
A particularly important constraint is a sufficiently large size of the
computational domain (small $k_1$), which is needed to obtain the expected
$\eta^{4/9}$ scaling of the correlation length.
When this constraint is not obeyed, the scaling is closer to $\eta^{1/3}$.
The second constraint of a sufficiently large Nyquist wave number is
important to obtain the correct values of the scale ratio of the positive
and negative magnetic helicity peaks, i.e., $k_{\rm II}/k_{\rm I}$.
Somewhat surprisingly, this ratio scales inversely with the initial scale
separation between the scale of the magnetic field and the CPI scale.
Increased values of $\aver{\mu_{\rm M}^-}$, which characterize the
strength of the CPI, are obtained when $\sigma$ is small or $|\mu_{50}|$
is large and therefore the coupling to the CPI is more efficient.

In the absence of spin flipping, even the slightest initial imbalance
will amplify as the magnetic energy decays; see \App{Imbalanced}.
On long time scales, this eventually leads to a fully helical state,
although simulations of this are at present unable to demonstrate this
conclusively owing to the finite size of the computational domain.
Spin flipping is another mechanism that can produce an imbalance between
magnetic helicity and fermion chirality.
In any case, however, the finally available magnetic energy and helicity
densities are always limited by the finiteness of the initial total
chirality imbalance.
For $\eta < \eta_\mathrm{ACC}$, when the chiral magnetic effect is not effective 
at the peak scale, magnetic helicity conservation governs
the decay of magnetic energy and the Hosking integral does not play a role.

We also discussed the implications of our findings for the 
generation of the baryon asymmetry of the Universe, in particular,
the scenario of baryogenesis from helicity decay.
The final baryon asymmetry in this scenario is controlled by
a dimensionless yield parameter $\chi$ that quantifies the helicity
density produced in the very early Universe, for instance, during a
stage of axion inflation.
In previous work, it was shown how the observed baryon asymmetry
can be generated from helicity decay at the time of the electroweak
phase transition for a specific $\chi$ value, $\chi_0 \sim 10^{-7}$; see Eq.~\eqref{eq:etaold}
and Ref.~\cite{Domcke:2022kfs}.
The situation at larger $\chi$ values, however, remained unclear.
At $\chi \gg 1$, one may have anticipated either (A) the overproduction
of baryon number or (B) catastrophic helicity erasure by the chiral
plasma instability and consequently no baryon asymmetry at all.

Thanks to the analysis in this paper, we now extend the previous analysis to the case where the decay of the hypermagnetic helicity occurs before the spin flipping for electrons becomes efficient, i.e., $\chi \gtrsim 10^{-4}$.
For typical parameters for axion inflation, although we initially have $|\mu_5| \ll k_\text{rh} / a_\text{rh}$, the opposite large hierarchy is eventually realized $|\mu_5| \gg k_\text{I} / a_\text{I}$.
Therefore the required value of $\chi$ to reproduce the observed baryon asymmetry depends on the value of the scaling index $q_5$ in the large hierarchy regime.
Further studies for the large hierarchy are indispensable for
understanding the outcome of baryogenesis.

\vspace{2mm}
{\bf Data availability}---The source code used for the
simulations of this study, the {\sc Pencil Code},
is freely available from Ref.~\cite{JOSS}.
The simulation setups and the corresponding data
are freely available from Ref.~\cite{SMZeroTotalChirality}.

\medskip
\begin{acknowledgments}
We thank V.~Domcke for fruitful discussions and J.~Warnecke for his work
on the implementation of SLD, which is used in some of the simulations.
We are also indebted to the referee for valuable suggestions.
Support through grant 2019-04234 from the Swedish Research Council (Vetenskapsr{\aa}det)
and NASA ATP award 80NSSC22K0825 (AB),
Grant-in-Aid for Scientific Research No.\ (C) JP19K03842 from the JSPS KAKENHI (KK),
MEXT Leading Initiative for Excellent Young Researchers  No.\ JPMXS0320200430 (KM),
Grant-in-Aid for Young Scientists No.\ JP22K14044
from the JSPS KAKENHI (KM),
and the grant 185863 from the Swiss National Science Foundation (JS)
are gratefully acknowledged.
We acknowledge the allocation of computing resources provided by the
Swedish National Infrastructure for Computing (SNIC)
at the PDC Center for High Performance Computing Stockholm and Link\"oping.
\end{acknowledgments}

\appendix

\section{Additional time and length scales}
\label{AdditionalScales}

In \Sec{IIE} and also later in this paper, we defined a number of
time scales and wave numbers.
In \Tab{TAnalyticTimeScale}, we summarized the characteristic 
time scales relevant for the evolution of the system.
In \Tab{TAnalyticTimeScale2}, we present a summary of additional time scales
and in \Tab{TWaveNumbers} various wave numbers defined in this paper.

\begin{table*}[t]\caption{
Summary of additional time scales used in this paper and not yet defined
in \Tab{TAnalyticTimeScale}.
}\vspace{12pt}\centerline{\begin{tabular}{llll}
Time scale & Definition & Explanation \\
\hline
$\eta_{\rm flip}$         & \Sec{DiagnosticQuantities}, \Eq{etaflip} & time when spin flipping is turned on \\
$\eta_{\rm off}$          & \Sec{DiagnosticQuantities}, \Eq{etaflip} & time when spin flipping is later turned off again \\
$\eta_0$                  & \Sec{InitialConditions}    & initial time \\
$\eta_{\rm5dec}$          & \Sec{IIE}, \Eqs{eta5dec}{eta5dec-SM}     & time when decay is determined by conservation of adapted Hosking integral \\
$\eta_\mathrm{I}$         & \Sec{EvolCharScales}                     & time when large-scale spectrum starts to decrease via inverse cascade \\
$\eta_\pm^\mathrm{(i)}$   & \Sec{EvolCharScales}                     & time when negative helicity modes at secondary peak starts to decrease \\
$\eta_\pm^\mathrm{(ii)}$  & \Sec{EvolCharScales}                     & time when decay of the secondary peak becomes slower with a smaller index \\
$\eta_{\mu_\mathrm{M}^+}$ & \Sec{subsec:evolution}                   & time when the ACC commences exhibiting a power law decay \\
$\eta_{\mu_\mathrm{M}^-}$ & \Sec{subsec:evolution}                   & time when the ACC decays again \\
$\eta_B^0$                & \Sec{Baryogenesis}, \Eqs{eq:etaold}{eq:etanew} & theoretical time when present-day baryon asymmetry is established \\
$\eta_B^{\rm obs}$        & \Sec{Baryogenesis}         & observed time when present-day baryon asymmetry is established \\
$\eta_\mathrm{sf}$        & \Sec{Baryogenesis}         & spin flipping time \\
\label{TAnalyticTimeScale2}\end{tabular}}\end{table*}

\begin{table*}[t]\caption{
Summary of wave numbers defined in this paper.
}\vspace{12pt}\centerline{\begin{tabular}{llll}
Wave number & Definition & Explanation \\
\hline
$\kp$        & \Sec{Intro} & equivalent to $\xiM^{-1}$, i.e., the time-dependent peak wave number of $\EM(k)$ \\
$k_1$        & \Sec{ModelDescription} & lowest wave number in the domain $=2\pi/L$ \\
$k_{\rm Ny}$ & \Sec{ModelDescription} & Nyquist wave number $=N\pi/L$ \\
$k_0$        & \Sec{DiagnosticQuantities}, \Eq{k0def} & wave number of the initial peak of $\EM(k)$ \\
$k_{\rm I}$  & \Sec{DiagnosticQuantities} & peak wave number of the positive peak of $\HM(k)$ \\
$k_{\rm II}$ & \Sec{DiagnosticQuantities} & peak wave number of the negative peak of $\HM(k)$ \\
$k_\pm$      & \Sec{DiagnosticQuantities} & wave number where $\HM(k)$ changes sign \\
$\mu_{50}$   & \Sec{DiagnosticQuantities} & initial value of $\mu_5$, critical CPI wave number \\
$k_{\rm rh}$ & \Sec{Baryogenesis} & wave number of the Hubble horizon at end of reheating \\
$k_{\rm sf}$ & \Sec{Baryogenesis} & inverse spin flipping time, i.e., $1/\eta_{\rm sf}$ \\
\label{TWaveNumbers}\end{tabular}}\end{table*}

\section{Other cases with spin flipping}
\label{OthersSpinFlip}

\begin{figure}\begin{center}
\includegraphics[width=\columnwidth]{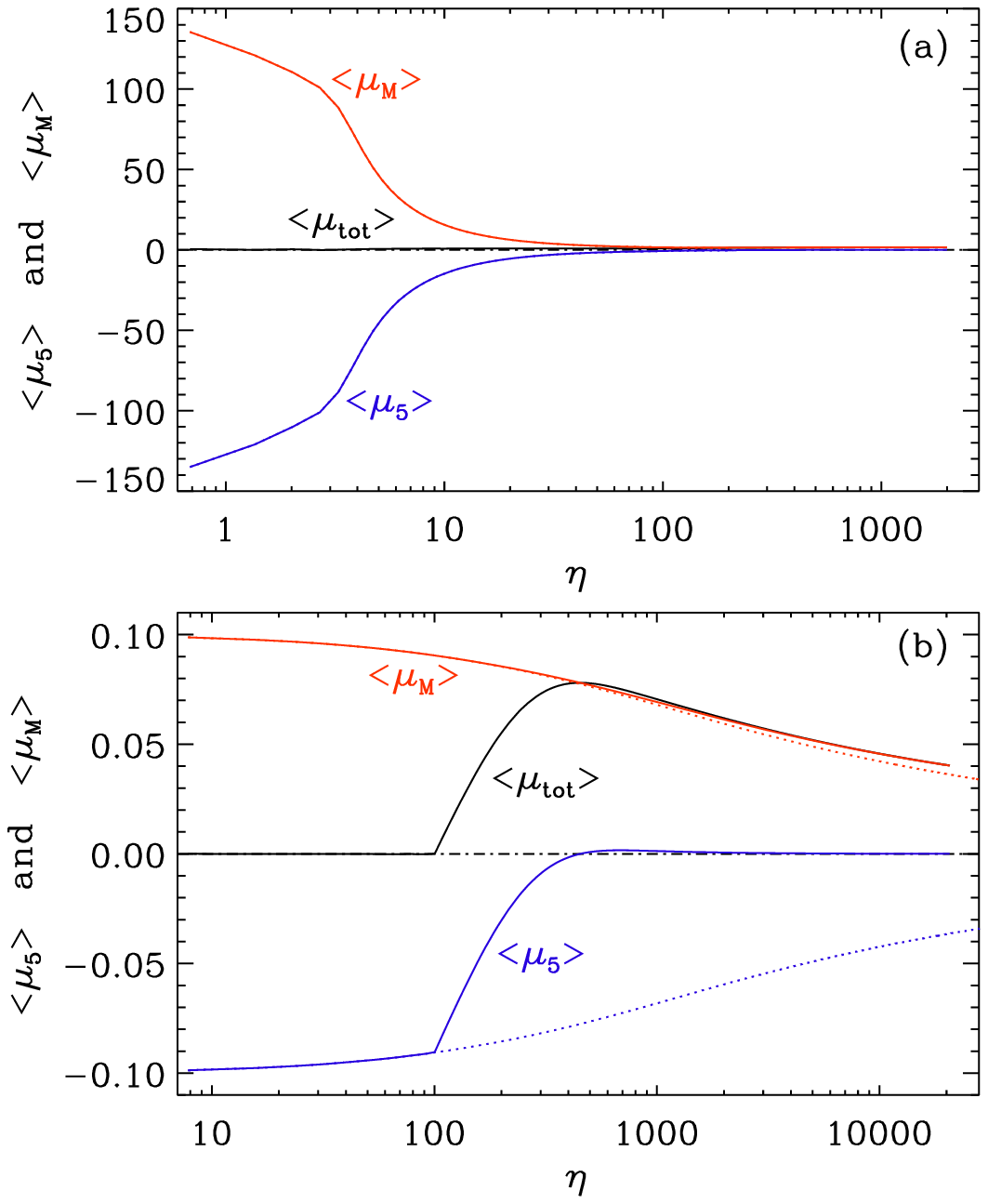}
\end{center}\caption[]{
Similar to \Fig{p_comp2}, but with parameters similar to Run~VI (a)
and Run~P (b), where $\eta_{\rm ACC}\approx13$ is much less than
$\eta_{\rm flip}$ and $\eta_{\rm ACC}\approx5\times10^4$ is much
greater than $\eta_{\rm flip}$.
As for Run~F in \Fig{p_comp2}, we have $\eta_{\rm flip}=100$ also here
for both runs, but do not consider finite values of $\eta_{\rm off}$.
}\label{p_comp2_2panel}\end{figure}

In \Sec{sec:IIIG}, we presented Run~F with spin flipping, whose
other parameters were similar to those of Runs~N and N', where
$\eta_{\rm ACC}\approx(\eta_{\rm CPI}\eta_{\rm diff})^{1/2}=10^3$.
and $|\mu_{50}|/k_0=5$.
We now discuss a spin flipping version of Run~VI (Run~P), where
$\eta_{\rm ACC}=13$ ($5\times10^4$) is much less (much greater)
than the spin-flipping time, and where $|\mu_{50}|/k_0=160$
($|\mu_{50}|/k_0=0.1$) is much larger (smaller) than for Run~F.
The results are shown in \Figsp{p_comp2_2panel}{a}{b}.

As expected, when chirality flipping becomes effective much before the
onset of ACC, we hardly see its effect; see \Fig{p_comp2_2panel}(a).
Conversely, when mild ACC has already started by the time the chirality
flipping becomes effective, we do not see qualitative differences to
the original Run~P; see \Fig{p_comp2_2panel}(b).
Hence, the overall understanding of the effect of chirality flipping
does not change, and the basic features of chirality flipping are best
captured in \Fig{p_comp2}.

%FIG18

\section{Behavior in imbalanced chirality decay}
\label{Imbalanced}

In \Sec{Conclusions}, we emphasized that even the slightest initial
imbalance between magnetic helicity and fermion chirality will amplify
as the magnetic energy decays.
It is therefore important to remember that the dynamics discussed in
this paper is specific to the case of balanced chirality, which is
arguably also the most generic case.
We know that the decay of magnetic energy and the increase of
the correlation length follow a different behavior in the completely
imbalanced case compared to the unbalanced one.
We now discuss the behavior for the mildly imbalanced case.
Here, we show that there is a tendency for the system to approach the
behavior of a completely imbalanced one.

We discuss two runs, Run~A where the initial $\aver{\muM}$ is enhanced
by 20\% compared with $|\aver{\mu_5}|$, and Run~B where it is decreased
by 20\%.
Apart from that, the runs are the same as Run~O, i.e., the run discussed
in Ref.~\cite{BKS23}.

%FIG19
\begin{figure}[t]\begin{center}
\includegraphics[width=\columnwidth]{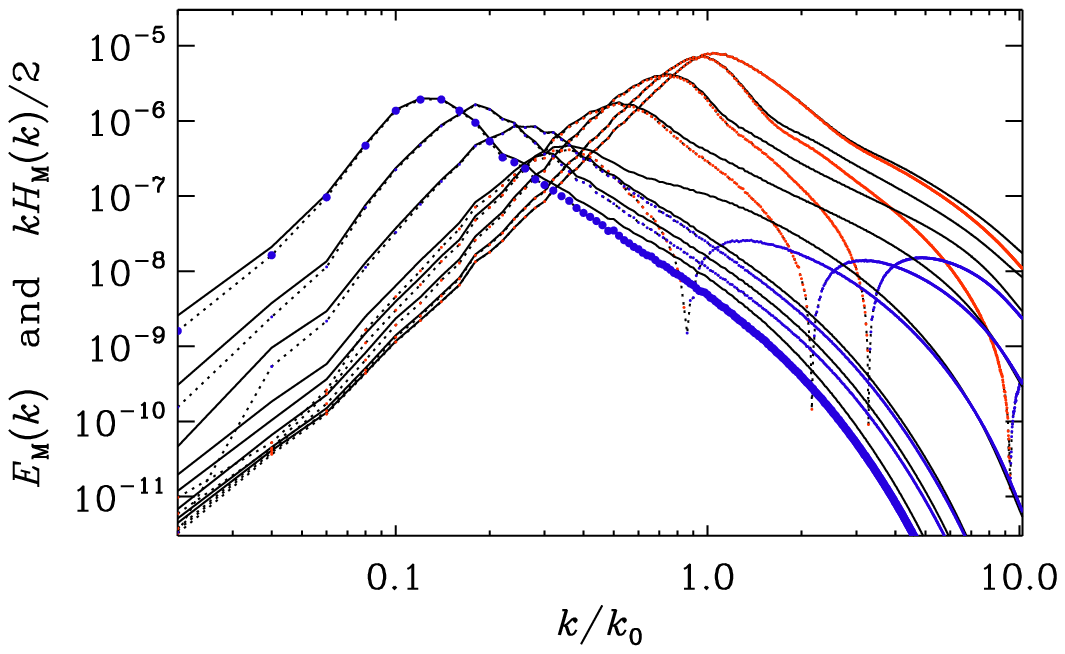}
\end{center}\caption[]{
Magnetic energy (solid lines) and normalized helicity spectra
$k\HM(k)/2$ (dotted lines with red and blue symbols for positive
and negative helicity spectra, respectively) for Run~A at
times $\eta=32$, 100, $320$, 1000, 3200, 15,000, 22,000, and 32,000.
}\label{rspec_select_1024a_mu10_k002o_less}\end{figure}

%FIG20
\begin{figure}[t]\begin{center}
\includegraphics[width=\columnwidth]{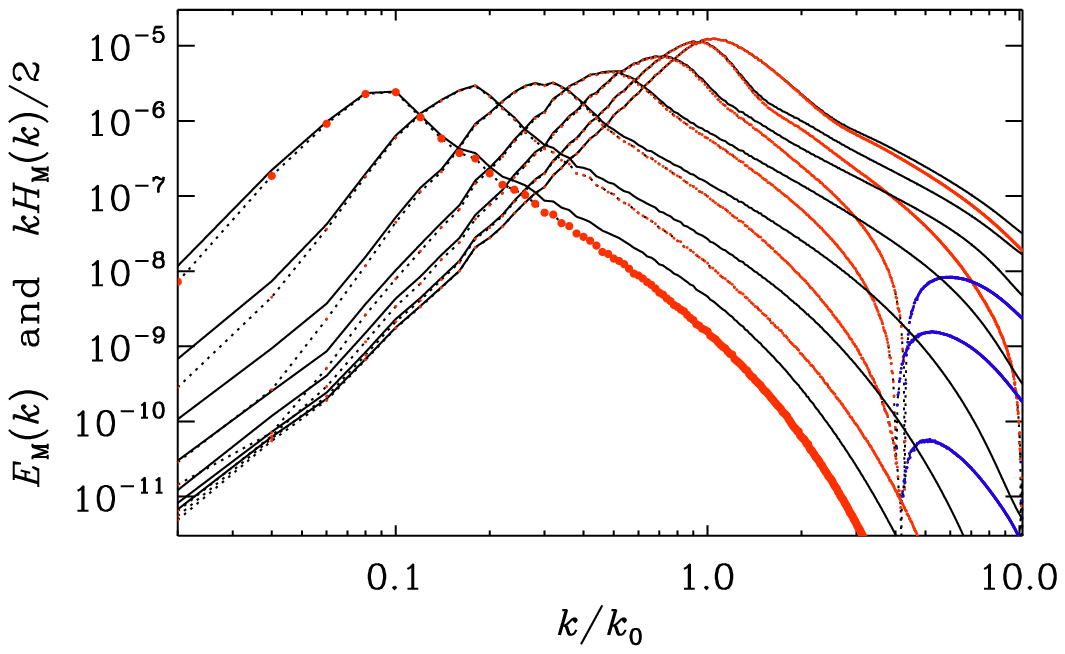}
\end{center}\caption[]{
Same as \Fig{rspec_select_1024a_mu10_k002o}, but for Run~B
at times $\eta=32$, 100, $320$, 1000, 3200, 10,000, and 32,000.
}\label{rspec_select_1024a_mu10_k002o_more}\end{figure}

%FIG21
\begin{figure}[t]\begin{center}
\includegraphics[width=\columnwidth]{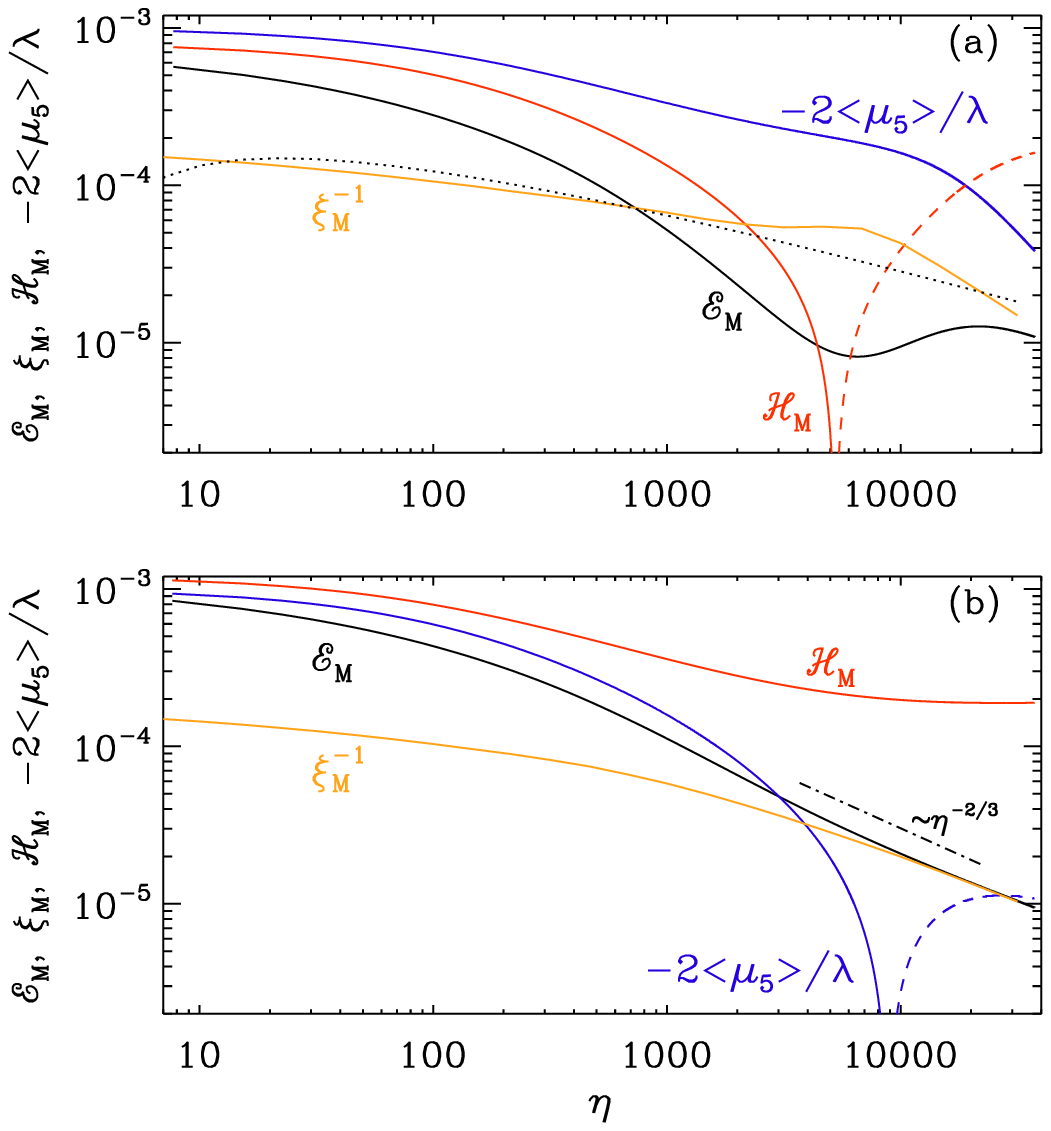}
\end{center}\caption[]{
Time dependence of $\EEM$ (black), $\xiM^{-1}$ (orange), $\HHM$ (red),
and $-2\aver{\mu_5}/\lambda$ (blue), for (a) Run~A with smaller and (b) Run~B
with larger magnetic helicity than in the balanced case.
Dashed lines indicate negative values; at late times
$-2\aver{\mu_5}/\lambda$ changes sign in (a), and
$\HHM$ changes sign in (b).
In (a) the dotted line denotes the $\eta^{-1/2}\log(\eta/\eta_{\log})$ scaling
of Ref.~\cite{Schober+2020} with $\eta_{\log}=3$.
}\label{penerg_comp}\end{figure}

In Run~A, where the magnetic helicity is weaker than in Run~O, the CPI
becomes dominant and overcompensates the magnetic helicity.
The net chirality is then negative.
Eventually, the sign of the magnetic helicity changes and all the
remaining fermion chirality is converted to magnetic fields with negative
helicity; see \Fig{rspec_select_1024a_mu10_k002o_less}, where we show
the magnetic energy $\EEM(k,t)$ and the normalized magnetic helicity
spectra $k\HM(k)/2$ for Run~A at times $\eta=32$, $320$, 1000, 3200,
10,000, and 32,000.
We see that $k|\HM(k)|/2$ approaches $\EEM(k)$ near the maximum.
In view of the spectral realizability condition, \Eq{realizability},
this means that the magnetic field is fully helical.
Away from the maxima, the inequality is no longer saturated, but this
is a typical effect in all turbulent flows where the current helicity
spectrum shows a Kolmogorov-type spectrum, making the magnetic helicity
spectrum therefore steeper than what could still be allowed by the
spectral realizability condition \cite{BS05b}.

On the other hand, when the fermion chirality is weak (Run~B),
the usual inverse magnetic cascade quickly gets established;
see \Fig{rspec_select_1024a_mu10_k002o_more}.
In either case, the fermion chirality gets ultimately
converted into magnetic helicity.
It just takes a little longer than when the magnetic helicity is
initially weak.
At the end, however, the usual inverse cascade for a fully helical
magnetic field commences.
The sign of magnetic helicity can be positive or negative, depending
on the sign of the initial total chirality.

To illustrate how the decay laws change when magnetic helicity and
fermion chirality no longer balance, we plot in \Fig{penerg_comp} the
time dependencies of $\EEM$, $\xiM$, $\HHM$, and $-2\aver{\mu_5}/\lambda$,
for (a) Run~A with 20\% smaller and (b) Run~B with 20\% larger magnetic
helicity than in the balanced case.
In both cases, we see a tendency of the decays of $\EEM$ and $\xiM^{-1}$
to slow down while those of $\HHM$ and $-2\aver{\mu_5}/\lambda$ follow
separate evolutions.
Especially in the case of Run~B, where the magnetic helicity
dominates of the fermion chirality, we see a tendency
toward a $\EEM\propto\xi^{-1}\propto t^{-2/3}$ as well as
$\mathcal{H}_\mathrm{M}=\const$ evolution, as expected from
magnetic helicity conservation.

\bibliography{ref}{}
\end{document}